\def\simlt{\lower.5ex\hbox{$\; \buildrel < \over \sim \;$}}
\def\simgt{\lower.5ex\hbox{$\; \buildrel > \over \sim \;$}}
\RequirePackage{lineno} 
 \documentclass[preprint2]{emulateapj}

\newcommand{\myemail}{mrl@gps.caltech.edu}
\slugcomment{Accepted to the Astrophysical Journal }
\shorttitle{Secondary Eclipse Atmospheric Retrieval}
\shortauthors{Line et al.}

\begin{document}
\title{A Systematic Retrieval Analysis of Secondary Eclipse Spectra I: A Comparison of Atmospheric Retrieval Techniques  }
\author{Michael R. Line}
\affil{Division of Geological and Planetary Sciences, California Institute of Technology, Pasadena, CA 91125}
\author{Aaron S. Wolf}
\affil{Division of Geological and Planetary Sciences, California Institute of Technology, Pasadena, CA 91125}
\author{Xi Zhang}
\affil{Division of Geological and Planetary Sciences,California Institute of Technology, Pasadena, CA 91125}
\author{Heather Knutson}
\affil{Division of Geological and Planetary Sciences,California Institute of Technology, Pasadena, CA 91125}
\author{Joshua A. Kammer}
\affil{Division of Geological and Planetary Sciences, California Institute of Technology, Pasadena, CA 91125}
\author{Elias Ellison}
\affil{Flintridge Preparatory School, La Ca\~nada, CA 91011}
\author{Pieter Deroo}
\affil{Jet Propulsion Laboratory, California Institute of Technology}
\author{Dave Crisp}
\affil{Jet Propulsion Laboratory, California Institute of Technology}
\author{Yuk L. Yung}
\affil{Division of Geological and Planetary Sciences,California Institute of Technology, Pasadena, CA 91125}
\email{mrl@gps.caltech.edu}
\altaffiltext{1}{Correspondence to be directed to \myemail}
\begin{abstract}
Exoplanet atmosphere spectroscopy enables us to improve our understanding of exoplanets just as remote sensing in our own solar system has increased our understanding of the solar system bodies. The challenge is to quantitatively determine the range of temperatures and molecular abundances allowed by the data which is often difficult given the low information content of most exoplanet spectra which commonly leading to degeneracies in the interpretation.   A variety of spectral retrieval approaches have been applied to exoplanet spectra, but no previous investigations have sought to compare these approaches.   We compare three different retrieval methods: optimal estimation, differential evolution Markov chain Monte Carlo, and bootstrap Monte Carlo on a synthetic water dominated hot-Jupiter.   We discuss expectations of uncertainties in abundances and temperatures given current and potential future observations.   In general we find that the three approaches agree for high spectral resolution, high signal-to-noise data expected to come from potential future spaceborne missions, but disagree for low resolution, low signal-to-noise spectra representative of current observations. We also compare the results from a parameterized temperature profile versus a full classical Level-by-Level approach and discriminate in which situations each of these approaches is applicable.  Furthermore, we discuss the implications of our models for the inferred C to O ratios of exoplanetary atmospheres. Specifically we show that in the observational limit of a few photometric points, the retrieved C/O is biased towards values near solar and near one simply due to the assumption of uninformative priors.  
\end{abstract}


\section{Introduction}
Thermal emission spectra ($\sim$ 1-30 microns) of extrasolar planets can tell us about their atmospheric temperatures and compositions  (see e.g., Charbonneau et al. 2005;  Tinetti et al. 2007; 2010; Grillmair et al. 2007; 2008;  Swain et al. 2009a; 2009b; Madahusudhan \& Seager 2009;  Stevenson et al. 2011; Madhusudhan et al. 2011; Lee et al. 2012;  Line et al. 2012).  At the moment, these observations come in two types, broadband photometry mainly from the Spitzer Space Telescope (see e.g., Knutson et al. 2010) and ground-based instruments (Croll et al. 2010; Anderson et al. 2010; Gibson et al. 2010; Deming et al. 2012; Gillon et al. 2012 ), as well as higher resolution spectra such as Hubble Space Telescopes Wide Field Camera 3 (WFC3) (Berta et al. 2012; Swain et al. 2012; Deming et al. 2013 )  and Near Infrared Camera and Multi-Object Spectrometer (NICMOS)  (Swain et al. 2009a; 2009b; Tinetti et al. 2010; Gibson et al. 2011; Crouzet et al. 2012).  From these observations, signatures of a variety of molecules have been detected including H$_2$O, CH$_4$, CO and CO$_2$ (Tinetti et al. 2007; Swain et al. 2009a; 2009b; Tinetti et al. 2010), although the robustness of some of these detections have recently been called into question (Gibson et al. 2011).  These same data have been used to infer the presence of atmospheric temperature inversions for a subset of hot Jupiters (e.g., Burrows et al. 2007, Knutson et al. 2008;2010; Forntey et al. 2008; Madhusudhan \& Seager 2009; 2010).   


While the above studies have given us insight into the nature of these planetary atmospheres, very few have focused on the uncertainties in temperatures and compositions.  Until relatively recently (Madhusudhan \& Seager 2009, Madhusudhan et al. 2011, Lee et al. 2012, and Line et al. 2012),  most compositions and temperatures and thus the subsequent conclusions, were determined through self-consistent forward modeling approaches that only explore a few potential solutions without a well-defined characterization of the uncertainty distributions of the physical parameters (e.g., Burrows et al. 2005; Fortney et al. 2005; Burrows et al. 2007).  Furthermore, some self-consistent solutions make physical assumptions that may not necessarily be valid in exoplanetary atmospheres such as the assumption of thermochemical equilibrium gas concentrations or radiative-convective temperature structures (that is they may ignore other potentially important processes such as vertical mixing, photochemistry, zonal winds etc.).    Additionally, this forward modeling approach is often unguided by the data and primarily driven by preconceived notions of how the atmosphere ``should" look (as pointed out by Lee et al. 2012 and Benneke \& Seager 2012) with the best solutions being the few that provide the lowest values of chi-squared.    

In order to more rigorously characterize the ranges of allowable temperatures and compositions, Madhusudhan \& Seager (2009) developed a multidimensional grid search approach which can fully characterize the uncertainty distributions for each parameter.  Subsequent studies (Madhusudhan et al. 2011; Benneke \& Seager 2012) used the more sophisticated Markov chain Monte Carlo approach (MCMC) to accomplish this goal.  However, such approaches require the computation of many millions of models in order to fully characterize the parameter uncertainties which may not be feasible for more sophisticated forward models with many free parameters.   In order to remedy this problem Lee et al. (2012) and Line et al. (2012) used the much faster optimal estimation (e.g, Rodgers 2000) approach to estimate the error distributions of each parameter.  This approach is much faster due to the assumption that the parameter error distributions are Gaussian.  However, this Gaussian assumption may result in an incorrect estimate of the error distributions (Benneke \& Seager 2012).  

The goals of this paper are to first understand the composition and temperature uncertainty distributions for different degrees of observational quality, and second to understand how those derived uncertainty distributions differ between the two fundamental parameter estimation approaches, optimal estimation and MCMC.  This investigation represents the first direct comparison and synthesis of these retrieval approaches as applied to exoplanet atmospheres.   A secondary goal is to understand how the derived composition uncertainties propagate into the C/O uncertainty.   We accomplish these goals by comparing three different retrieval algorithms: optimal estimation (OE), a new MCMC algorithm known as differential evolution Markov chain Monte Carlo (DEMC), and the model-dependent bootstrap Monte-Carlo approach (BMC).  This investigation is analogous to the investigation carried out by Ford (2005) on radial velocity data.   First we will describe the three different retrieval techniques as well as our forward model in \S \ref{sec:methods}.  We call our three-pronged retrieval approach CHIMERA-{\bf C}altec{\bf H} {\bf I}nverse {\bf M}od{\bf E}ling and {\bf R}etrieval {\bf A}lgorithms.     Second, we compare the three spectral retrieval methods on different synthetic spectral data sets of varying observational quality in order to assess the robustness of the error estimations from each approach in \S \ref{sec:three}.  We will also compare the parameterized temperature profile approach (e.g., Madhusudhan \& Seager 2009; Line et al. 2012) with the Level-by-Level profile approach (Lee et al. 2012).    Finally, we will discuss the implications of these uncertainties for the estimated C to O ratios.   

\section{Methods}\label{sec:methods}
In this section we describe the retrieval techniques, the forward model, and the parameterizations we use to retrieve the temperatures and compositions from thermal emission spectra.

\subsection{The Retrieval Techniques}
We use three different retrieval techniques to infer the compositions and temperatures from a spectrum.   The techniques are inherently Bayesian as they attempt to solve the inverse problem by summarizing the full shape of the posterior in terms of the location in parameter space of the maximum likelihood and the uncertainties about that location.  The first, and the fastest (least number of forward model calls) of these approaches, is optimal estimation, the second is the model-dependent bootstrap Monte Carlo,  and the third is differential evolution Markov chain Monte Carlo.   

\subsubsection{Optimal Estimation (OE)}\label{sec:OE}
The optimal estimation retrieval approach is well established in the fields of Earth atmosphere remote sensing  (Rodgers 1976; Towmey 1996; Rodgers 2000;  Livesay et al. 2006; Kuai et al. 2013), solar system atmosphere remote sensing (Conrath et al. 1998; Irwin et al. 2008; Nixon et al. 2007; Fletcher et al. 2007; Greathouse et al. 2011), and recently exoplanet atmosphere remote sensing (Lee et al. 2012; Line et al. 2012).  The basic approach is to minimize a cost function to obtain the {\em maximum a posteriori} (MAP) solution.  Using Bayes theorem and the assumption that the data likelihood and the prior are Gaussian , one can derive the following cost function (or log likelihood):  
\begin{eqnarray} \label{eq:cost_func}
J({\bf x})={({\bf y}-{\bf F(x)})^{T}{\bf S_{e}^{-1}}({\bf y}-{\bf F(x)})} \nonumber \\
+({\bf x}-{\bf x_{a}})^{T}{\bf S_{a}^{-1}}({\bf x}-{\bf x_{a}}) 
\end{eqnarray}
where $\bf y$ is the set of $n$ observations, $\bf x$ is the set of $m$ parameters which we wish to retrieve or the state vector, $\bf F(x)$ is the forward model that maps the state vector onto the observations (described in \S \ref{sec:forward_model}), and $\bf S_{e}$ is the $n \times n$ data error covariance matrix (typically off diagonal terms are zero and the diagonal elements are the square of the 1$\sigma$ errors of the observations).   $\bf x_{a}$ is the {\em a priori} state vector and $\bf S_{a}$ is the $m \times m $ {\em a priori} covariance matrix.    The first term in equation \ref{eq:cost_func} is simply ``chi-squared" and the second term represents the prior knowledge of the parameter distribution before we make the observations.  For high quality observations the second term is generally not important as most of the information in constraining the state vector comes from the observations.  For low quality observations it is just the opposite.  Following Irwin et al. 2008 we minimize equation \ref{eq:cost_func} with Newton's iteration method given by
\begin{eqnarray}
\bf x_{i+1}= x_{a}+S_{a}^{-1}K_{i}^{T}(K_{i}S_{a}^{-1}K_{i}^{T}+S_{e}^{-1})\nonumber\\
\bf (F(x)-y-K_{i}(x_{a}-x_i))
\end{eqnarray}
where $i$ is the iteration index and $\bf K_i$ is the Jacobian matrix at $i$ ($K_{nm}=\frac{\partial F_{n}}{\partial x_{m}}$).  Rather than taking the full Newton step we damp the solution with
\begin{equation} \label{eq:damp}
\bf x_{i+1}^{'}=x_{i}+\frac{x_{i+1}-x_{i}}{1+\zeta}
\end{equation}
where $\zeta$ is the damping parameter. At each iteration we evaluate $\bf J(x_{i+1})$ and $\bf J(x_{i+1}^{'})$.  If the latter is smaller, we set the state vector for the next iteration to  $\bf x_{i+1}^{'}$ and decrease $
\zeta$ by 0.3.  Otherwise we keep increasing $\zeta$ by a factor of 10 and re-evaluate equation \ref{eq:damp} and equation \ref{eq:cost_func} until $\bf J(x_{i+1}^{'})$ becomes less than $\bf J(x_{i+1})$.   Convergence is achieved when $\bf J$  changes by less than $1\times10^{-6}$ from the previous iteration, which typically occurs after $\sim$10s of iterations.  The resulting state vector is the MAP solution, or the ``best-fit".  Assuming that the posterior is normal which is achieved by linearizing the forward model about the best-fit solution, the uncertainties on the state vector parameters are given by the posterior covariance matrix
\begin{equation}  \label{eq:covariance_matrix}
{\bf \hat{S}}={\bf (K^TS_e^{-1}K+S_a^{-1})^{-1}}
\end{equation}
Again, this matrix represents a multi-dimensional {\em normal} distribution (see Rodgers 2000 for the derivation). The diagonal elements are the square of the marginalized errors whereas the off diagonal terms describe the correlations/degenracies amongst the parameters.  The first term ,$\bf K^TS_e^{-1}K$, represents the uncertainties due to the measurement errors.  This term uses the local gradient information to estimate the parameter uncertainties.   The second term represents the prior uncertainties before making the measurements, which has less influence for higher quality data.  Again, the major assumption in equations  \ref{eq:cost_func} and \ref{eq:covariance_matrix} is that the posterior for each paramter is Gaussian.  This assumption is only valid when the region in phase space over which the forward model can be linearized is broader than the parameter uncertainties.  However, it is this assumption that allows this approach to be extremely fast requiring only tens of forward model calls, which given the speed of our forward model ($\sim$5 seconds), results in a full retrieval in only a few minutes.    As we shall see in $\S$\ref{sec:three}, this assumption is valid for data that is of high resolution and signal-to-noise, but breaks down for low resolution, low signal-to-noise data.  In order to ensure that the global minimum of equation \ref{eq:cost_func} is found, multiple starting guesses are used.  They generally all converge to the same solution.

\subsubsection{Model Dependent Bootstrap Monte-Carlo (BMC)}\label{sec:BMC}	
A common way to more robustly characterize errors is through a Monte-Carlo resampling of the data (see e.g., Press et al. 2002 Chapter 15.6, Ford 2005 $\S$4.2) in which many thousands of realizations of the original data (in our case, the spectra) are created using the uncertainties from the original dataset.  These synthetic data are then refit using, say, OE, and the resulting best fit parameter distributions represent the uncertainties.  There are multiple ways of generating the synthetic data realizations.  The most common way is the residual resampling approach in which data realizations are created by adding the randomly shuffled residual between the best fit model and the data back to the original best fit model.    This new realization is then fit and the process is repeated many times.  The approach we take is similar, but rather than generate a new spectrum using the residual, we simply take the best fit, from OE, and then resample each point by drawing it from a normal distribution with a mean given by the best fit value and the width given by the data error bar for that point.   We chose this approach over the residual resampling approach because sparse coverage spectra, like those from broadband observations, have virtually no residual as they can be fit perfectly due to the greater number of parameters than data points.  We typically generate $\sim$1000 spectra realizations that are then refit by optimal estimation to obtain the state vector parameter distributions.  

\subsubsection{Differential Evolution Markov Chain Monte Carlo (DEMC)}\label{sec:DEMC}
The MCMC approach has revolutionized parameter estimation and error analysis in many fields.  It is routinely used in radial velocity (Ford 2005 ) and transit light curve (e.g., Eastman, Gaudi, \& Agol 2013) error analysis.  Results from a well converged MCMC analysis can generally be considered as the best possible representation of the parameter uncertainties. Recently, this approach has been applied to the exoplanet atmosphere retrieval problem (Madhusudhan et al. 2011; Benneke \& Seager 2012).  Unlike optimal estimation, MCMC approaches make no assumptions about the shape of the posterior, but rather evaluate the posterior with millions of samples.

The basic approach of MCMC is to sample the posterior through a random walk process.   The random walk is carried out  by drawing states from some proposal distribution and evaluating whether or not the proposed state has an increased likelihood over the previous.  Typically the proposal distribution is a normal distribution with a mean given by the current state in the chain ($\bf x_i$) and a user defined width to achieve a particular acceptance rate (Gibbs sampling or Metropolis-Hastings).   If the proposed state ($\bf x_p$) has an improved likelihood over the current state, then that state is kept ($\bf x_{i+1}=x_p$) and a new proposal is made from that location.  If the proposal state has not improved the likelihood, then that state is either rejected or accepted with some probability.     This previous state dependent random walk constitutes a Markov chain.  Given enough samples this Makov chain will converge to the target posterior (see Ford 2005 for a more detailed explanation).  

Rather than standard MCMC approaches, we use an adaptive algorithm known as differential evolution Markov chain Monte Carlo (DEMC) (ter Braak 2006; ter Braak \& Vrugt  2008).    The purpose of this approach is to obtain more appropriate proposal states by identifying the proper scale and orientation of the current estimate of the posterior.  This scale and orientation information comes from the chain history.  This approach gives a more efficient probing method for highly correlated parameter spaces and yields improved convergence rates.  Our DEMC procedure is as follows:\\

1.  Apply the OE technique to the measurements to obtain the best fit state vector and posterior covariance matrix,  $\bf \hat S$.  This step provides an initial estimate of the posterior.\\

2. Initialize $N_{init}$ links ($\bf x_{i=0-N_{init}}$) in each of $N_{chains}$ (typically 3 chains, more chains will slow convergence) independent chains (arrays) by randomly drawing state vectors from the multivariate normal described by the posterior covariance matrix from step 1.  Set the last link in one of the chains to the best fit state vector obtained in step 1.    This step provides a good starting history from which our initial proposal states can be drawn.  Combine each of the independent chains into one long chain that composes the history, $\bf X_{history}$.  \\

3. Evaluate the cost function, $\bf J$, in equation \ref{eq:cost_func} for the last link in each of the chains.   If using a flat prior ignore the second term.  This, again, is simply the equivalent of evaluating chi-squared.  \\

4.  Draw two random numbers, $R_1$ and $R_2$, between zero and $N_{chains}\times i$, where $i$ is the current state in the chain.  Initially, $i=N_{init}$.   Evaluate the proposed jump state given by
\begin{equation}
\bf x_{p}=x_{i}+\gamma(x_{R1}-x_{R2})+e
\end{equation}
where $\bf x_{R_1}$ and $\bf x_{R_2}$ are the states from different points in the the chain history, $\bf X_{history}$.  $\gamma$ is a scale factor typically set to $2.38/\sqrt(2*m)$ (ter Braak 2006), where $m$ is the number of parameters.  This factor is meant to give acceptance rates of $\sim$0.23 for large $m$.   $\bf e$ is a vector drawn from a multivariate normal distribution with a small variance relative to the chain variance in order to introduce a small amount of additional randomness.  Repeat this process for the other $N_{chains}-1$ chains.\\

5.  Evaluate the Metropolis (Metropolis et al. 1953) ratio, $r=P({\bf x_{p}})/P({\bf x_{i}})=e^{-\frac{1}{2}(\bf J(x_{p})-J(x_{i}))}$.   If $r$ is larger than 1, set $\bf x_{i+1}=x_p$ and if it is smaller only accept if it is larger than a random number between 1 and 0.  Otherwise do not update the chain, set  $\bf x_{i+1}=x_i$.  Repeat for the other  $N_{chains}-1$ chains.  Add the updated links in all $N_{chains}$ to $\bf X_{history}$.\\

6. Repeat steps 4 and 5 until convergence is met.  Convergence can be determined by looking at the trace plots of $\bf X_{history}$ for each parameter or by using the Gelman-Rubin statistic on the set of $N_{chains}$ chains.  For this we use the algorithm from Eastman, Gaudi, \& Agol 2013 which requires the Gelman-Rubin statistic to be less than 1.01 and the number of independent draws to be greater than 1000 for each parameter .  Convergence typically occurs in less than $10^{5}$ links in each of the $N_{chains}$ for a total of $N_{chains}\times10^{5}$ links, ***with given the speed of our forward model takes $\sim$5 days for a typical run.  This is about an order of magnitude less than parallel tempering or pure Metropolis-Hastings.

 \subsection{The Forward Model}\label{sec:forward_model}
The forward model, $\bf F(x)$, is the most important part of any retrieval algorithm.  It is what maps the state vector of retrievable parameters onto the observations.   In the case of atmospheric retrieval, the forward model takes temperatures and compositions and generates a model spectrum.      Our particular forward model numerically solves the planet-parallel thermal infrared radiation problem for an absorbing, emitting atmosphere (we neglect scattering).   We first divide the atmosphere into $N_{lev}$ discretized atmospheric layers.  The absorption optical depth for the $k^{th}$ gas in the $z^{th}$ layer at  wavelength $\lambda$ is
\begin{equation}
\Delta\tau_{k,z,\lambda}=f_{k,z}\sigma_{k,z,\lambda}\frac{\Delta P_{z}}{\mu_{atm}g}
\end{equation}
where  $f_{k,z}$ is the volume mixing ratio of the $k^{th}$ gas in the $z^{th}$ layer , $\sigma_{k,z,\lambda}$ is the absorption cross section per molecule of the  $k^{th}$ gas in $z^{th}$ layer at  wavelength $\lambda$, $\Delta P_{z}$ is pressure thickness of the $z^{th}$ slab, $\mu_{atm}$ is the mean molecular weight of the atmosphere and $g$ is the gravity.  The absorption cross sections are pre-computed on a 1 $cm^{-1}$ wavenumber grid at 20 evenly spaced temperature and log-pressure points from 500-3000K and 50-$10^{-6}$ bars respectively (similar to Sharp \& Burrows 2007).  The cross sections for each wavelength on the pre-computed grid are interpolated to the atmospheric temperatures and pressures in the $z^{th}$ slab.     To compute the total slab optical depth we sum the contribution from each gas to obtain
\begin{equation}
\Delta\tau_{z,\lambda}=\sum_{k=1}^{N_{gas}}\Delta\tau_{k,z,\lambda}
\end{equation}
Upon computing the optical depths at each level, we can now solve for the upwelling irradiance with
\begin{equation} \label{eq:RT_eq}
I_{\lambda}=\sum_{z=0}^{N_{lev}}B_{\lambda}(T_{z})e^{-\sum_{j=z}^{N_{lev}}\Delta \tau_{j,\lambda}}\Delta \tau_{z,\lambda}
\end{equation}
where $N_{lev}$ is the number of atmospheric levels and $B_{\lambda}(T_{z})$ is the Planck function at wavelength $\lambda$ and temperature in the $z^{th}$ slab.  We use 90 atmospheric layers to compute the upwelling flux.  

An important part of the forward model when using the optimal estimation approach is the computation of the Jacobian, or the sensitivity to the upwelling irradiance with respect to the desired retrievable parameters.     When possible, it is preferable that the Jacobian be calculated analytically for both improvements in speed and in accuracy.  We are interested in the retrieval of both abundances and temperatures so we must compute the Jacobian with respect to both the abundances and temperatures.  We make the assumption of vertically uniform gas mixing ratios throughout the atmosphere and hence, $f_{k,z}$ is independent of $z$.  We now differentiate equation \ref{eq:RT_eq} with respect to the uniform gas mixing ratios for each gas $f_k$ to obtain
\begin{eqnarray}\label{eq:gas_jac}
\frac{\partial I_{\lambda}}{\partial f_{k}}=\sum_{z=0}^{N_{lev}}B_{\lambda}(T_{z})e^{-\sum_{j=z}^{N_{lev}}\Delta \tau_{j,\lambda}}\frac{\Delta \tau_{k,z,\lambda}}{f_{k}}\nonumber\\
 - \sum_{z=0}^{N_{lev}}(B_{\lambda}(T_{z})e^{-\sum_{j=z}^{N_{lev}}\Delta \tau_{j,\lambda}}\Delta \tau_{z,\lambda}\sum_{j=z}^{N_{lev}}\frac{\Delta \tau_{k,j,\lambda}}{f_{k}})
\end{eqnarray}
The first term is due to the changing emissivity of the emitting slab and the second term is how the change in transmittance affects the upwelling irradiance.  

The sensitivity of the irradiance to a change in temperature in the $z^{th}$ slab is given by
\begin{eqnarray}\label{eq:T_jac}
\frac{\partial I_{\lambda}}{\partial T_{z}}=(e^{-\sum_{j=z+1}^{N_{lev}}\Delta \tau_{j,\lambda}}-e^{-\sum_{j=z}^{N_{lev}}\Delta \tau_{j,\lambda}})\frac{\partial B_{\lambda}(T_{z})}{\partial T_{z}}
\end{eqnarray}
This equation is similar to equation 14 in Irwin et al. 2008 but we have neglected the first and last terms in their formula as they are small.

Since the observations are reported as the ratio of the planet flux to the stellar flux and not the irradiance, we perform a disk integration of equations \ref{eq:RT_eq}-\ref{eq:T_jac}  using four point Guassian quadrature and then divide by an interpolated PHOENIX stellar flux grid model (Allard et al. 2000) .  

\begin{figure*}[h]
\begin{center}
\includegraphics[height=4.5in,width=!, angle=0]{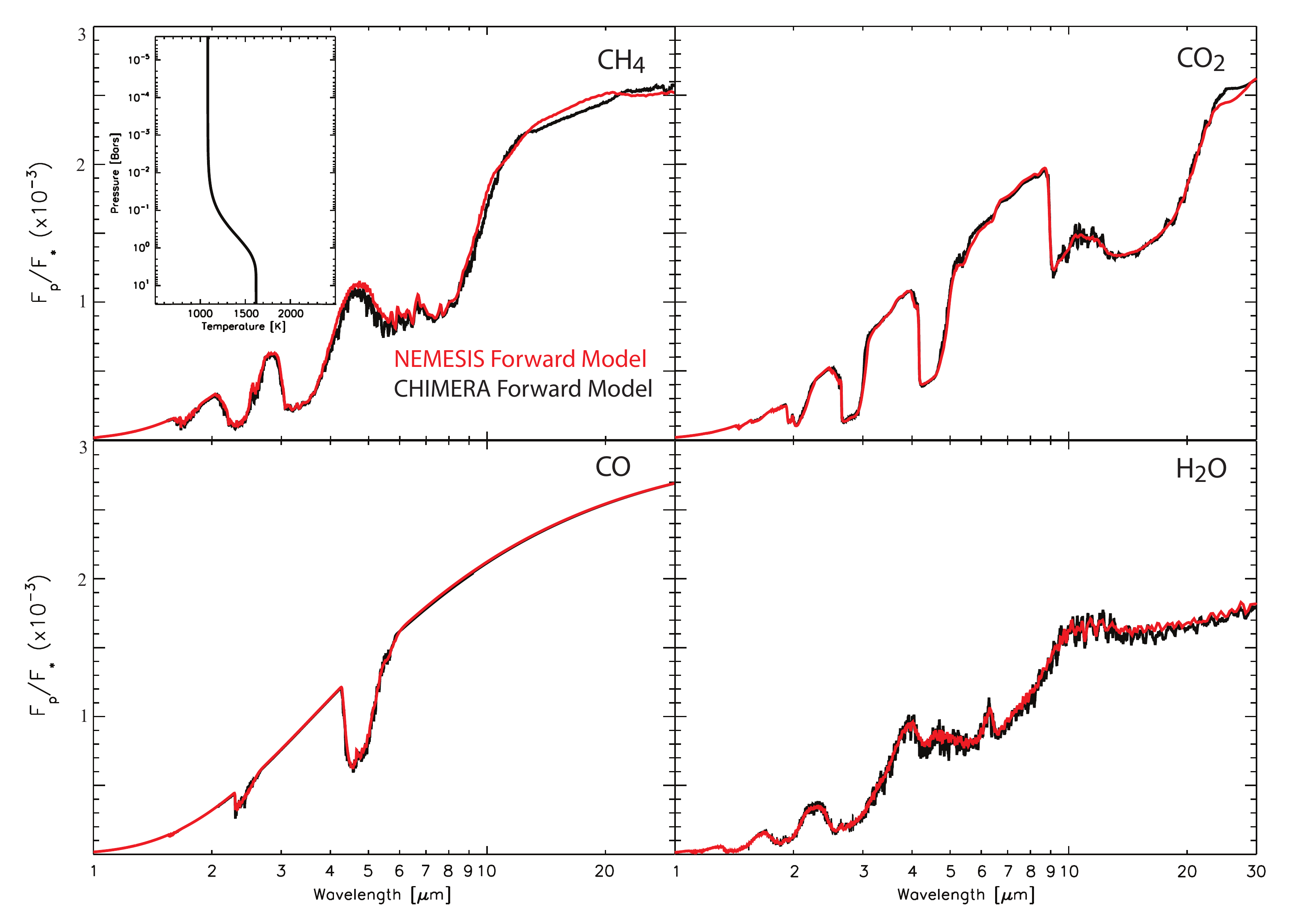}
\end{center}
\caption[Radiative Transfer Code Validation]{ \label{fig:figure1}Comparison of the thermal emission spectrum from our forward model (black) with the NEMESIS forward model (red).  The temperature-pressure profile is shown in the inset.  For this comparison we assume  uniform mixing ratios of $10^{-4}$ for CH$_4$ , CO$_2$, CO, and H$_2$O.  H$_2$ is set to 0.85 and He is set to 0.15.  This planet is assumed to be hydrogen dominated (mean molecular weight of 2.3 amu) with a radius of 1R$_{J}$, a gravity of 22 ms$^{-2}$ orbiting a 5700 K pure blackbody star with a radius of 1R$_{sun}$.   }
\end{figure*}

We include only CH$_4$, CO$_2$, CO, H$_2$O, H$_2$, and He in our model.   H$_2$ and He are fixed in our models at thermochemically justifiable abundances.  The exact abundances of these species is not critical as the sensitivity of the spectrum to H$_2$ and He is minimal.  We retrieve only CH$_4$, CO$_2$, CO, and H$_2$O.  We choose these species because they are the most spectroscopically active and abundant species.  Admittedly we could/should include every possible atmospheric constituent but this would be unwieldy and reliable high temperature absorption line lists only exist for a few.  On that note, we use the HITEMP database (Rothman et al. 2010) to compute the tabulated cross-sections for CO$_2$, CO, and H$_2$O and the STDS database for CH$_4$ (Wenger \& Champion 1998).  Below 1.7 $\mu m$ for CH$_4$  we simply use the HITRAN (Rothman et al. 2009) database for lack of anything better (to the best of our knowledge).  We use the Barysow et al. (2001;2002) and J{\o}rgensen et al. (2000)  databases for the computation of the H$_2$-H$_2$/He collision induced opacities.  The Reference Forward Model (RFM-http://www.atm.ox.ac.uk/RFM/) was used to compute the tabulated cross sections from the line strength databases.  We have validated our forward model through a detailed comparison with the Oxford NEMESIS group (e.g., Lee et al. 2012) and our results agree to better than 5$\%$ (see Figure \ref{fig:figure1}).

An additional component of the forward model is the instrumental function used to convolve with the high-resolution model spectrum.  For the broadband points we simply integrate the flux from the high-resolution model spectrum with the appropriate filter function for that point.  When fitting higher resolution observations, the instrumental function is assumed to be a Gaussian (valid for grating spectrometers) in wavelength with a full width at half maximum (FWHM) determined by observations.

Now that we have a well defined forward model we can define our state vector.  Again, we wish to retrieve the abundances of CH$_4$ , CO$_2$, CO, and H$_2$O and the temperature profile.  More specifically, we choose to retrieve the log of the abundances as they can vary by orders of magnitude and to prevent negative mixing ratios.  Our state vector is given by
\begin{equation}\label{eq:state_vec}
{\bf x}=[log(f_{H_2O}), log(f_{CH_4}), log(f_{CO}), log(f_{CO_2}), T]^T.
\end{equation}
where the $f_k$'s are all assumed constant with altitude.  We feel this is appropriate for two reasons.  First, vertical mixing will smooth out the mixing ratio profiles over the thermal infrared photosphere (Line et al. 2010, Moses et al. 2011, Line et al. 2011), and secondly current observations simply do not have the information content to warrant the retrieval of vertical mixing ratio information (see Lee et al. 2012).    In the next section we describe how to go about retrieving the temperature profile.

\subsubsection{Parameterized vs. Level-by-Level (Level-by-Level) Temperature Profile}\label{sec:TP_approach}\label{sec:par_vs_Level-by-Level}	
We employ two approaches to retrieve the temperature profiles.  The first, and the most commonly used in Earth and solar system atmosphere retrieval problems, is the Level-by-Level approach.  This is the approach used in Lee et al. 2012.  The second is a parameterized temperature profile approach similar to the approach used in Madhusudhan \& Seager 2009 and Line et al. 2012.  Each has its advantages and disadvantages described below.

The Level-by-Level temperature retrieval approach seeks an estimate of the temperature at each of the $N_{lev}$ model layers.   This approach is advantageous in that there are no pre-conceived assumptions made about how the atmospheric temperature should be parameterized.    If the spectral signal-to-noise and resolution are high enough, there is generally enough sensitivity to obtain information about the temperature in individual atmospheric layers.   However, there is a finite vertical resolution given the quality of the observations.  Typically this resolution is set by the width of the thermal emission weighting functions and how much they overlap.  Generally, when the spectra are noisy the Level-by-Level approach fits the noise which results in unphysical structure in the retrieved temperature profile.  This is analogous to fitting a high-degree polynomial to only a few points.   There are ways to smooth unphysical structure, one of them being to assume a correlation among the atmospheric layers (Rodgers 2000, Irwin 2008) implemented through the prior covariance matrix, $\bf S_{a}$, with 
\begin{equation}\label{eq:TP_covar}
S_{a,ij}=(S_{a,ii}S_{a,jj})^{1/2}e^{\frac{-|ln(P_i/P_j)|}{h}}.
\end{equation}
Here $P_i$ and $P_j$ are the pressures at the $i^{th}$ and $j^{th}$ levels, respectively and $h$ is the correlation length that controls the level of smoothing.  The correlation length can be thought of as the number of scale heights over which the temperatures are correlated.     For our simulations we choose $h=7$ as this provides a sufficient level of detail without producing unphysical oscillations.    When using this approach our state vector is exactly as it is in equation \ref{eq:state_vec} with $T$ being an $N_{lev}$ vector of temperatures at each level.  The Level-by-Level approach is only appropriate when the information content of the spectra is sufficiently high such that the addition of the $N_{lev}$ additional parameters is justified.  For most current exoplanet spectra, this is an invalid approach.

The second temperature profile retrieval approach makes use of a parameterization.  This approach is advantageous when the information content of a spectrum is low as the number of free variables is greatly reduced.  However, the parameterization does force the retrieved atmospheric temperature structure to conform only to the profile shapes and physical approximations allowed by that paramterization.  For our particular parameterization, we assume the atmosphere to be in radiative equilibrium based upon the analytic radiative equilibrium temperature profile of Guillot 2010 (and others such as Hansen 2008; Heng et al. 2012; Robinson \& Catling 2012).  This is the same parameterization used in Line et al. 2012.   This profile assumes two independent downwelling visible channels of radiation and one upwelling stream of thermal emission.  Briefly, the temperature as a function of the thermal optical depth,$\tau$, is given by
\begin{equation}\label{eq:TP_par}
T^4(\tau)=\frac{3T^4_{int}}{4}(\frac{2}{3}+\tau)+\frac{3T^4_{irr}}{4}(1-\alpha)\xi_{\gamma_{1}}(\tau)+\frac{3T^4_{irr}}{4}\alpha \xi_{\gamma_{2}}(\tau)
 \end{equation}
with
\begin{equation}
\xi_{\gamma_{i}}=\frac{2}{3}+\frac{2}{3\gamma_{i}}[1+(\frac{\gamma_{i}\tau}{2}-1)e^{-\gamma_{i}\tau}]+\frac{2\gamma_{i}}{3}(1-\frac{\tau^{2}}{2}){\rm E_{2}}(\gamma_{i}\tau)
\end{equation}
where $\gamma_{1}$ and  $\gamma_{2}$ are the ratios of the Planck mean opacities in the visible streams to the thermal stream and the parameter $\alpha$ (range 0 to 1) partitions the flux between the two visible streams.  $E_2(\gamma \tau)$ is the second order exponential integral function.  The internal heat flux is parameterized by the temperature ,T$_{int}$, which is fixed since it has little impact on the spectra.  The stellar input at the top of the atmosphere is represented by T$_{irr}$ given by
\begin{equation}
T_{irr}=\beta(\frac{R_{*}}{2a})^{1/2}T_{*}
\end{equation}
where $R_{*}$ and $T_{*}$ are the stellar radius and temperature, $a$ is the semi-major axis, and $\beta$ is a catch all term on the order of unity for the albedo, emissivity, and day-night redistribution.  The grey infrared optical depth can be mapped onto pressure coordinates using
\begin{equation}\label{eq:tau}
\tau=\frac{\kappa_{IR}P}{g}
\end{equation}
where $P$ is the pressure, $g$ the surface gravity (at 1 bar), and $\kappa_{IR}$ the Planck mean thermal infrared opacity.  This $\tau-P$ mapping assumes a linear relation between the optical depth and pressure, or a pressure independent infrared opacity.  More complicated mappings that account for the pressure dependence of $\kappa_{IR}$ can also be used (see e.g., Robinson \& Catling 2012).    

This temperature parameterization has 5 free parameters governing its structure: $\kappa_{IR}$, $\gamma_{v_{1}}$, $\gamma_{v_{2}}$,  $\beta$, and $\alpha$.   Our parameterized state vector again, is given by equation \ref{eq:state_vec} but with $T$ replaced with [$\kappa_{IR}$,  $\gamma_{v_{1}}$,  $\gamma_{v_{2}}$,  $\beta$, $\alpha$].  Combined with the gases this gives a total of 9 free parameters.  The temperature {\em profiles} are then reconstructed from the probability distributions of those 5 parameters.

We should note that currently most exoplanet spectra often have fewer measurements than desired state variables.  This means that each parameter cannot be uniquely determined.  This is not a new problem (e.g., Madhusudhan \& Seager 2009).  This is  why the prior is crucial.  We can think of the prior as an artificial set of ``data" on which the retrieval (all retrieval approaches) can rely when the measurements are insufficient to constrain a given parameter.  Therefore the resulting constraints on a given parameter are a combination of the information obtained from the spectra and the prior knowledge.  In the extreme case of no observational constraint, the posterior will simply be the prior.  Hence, it is critical to choose an appropriate prior, especially for cases when there are more parameters than measurments.  

 With the optimal estimation formalism, we can assess the degree to which the constraint comes from the measured spectra versus the prior through what is called the averaging kernel.  The averaging kernel is an $m \times m$ matrix whose elements are given by
 \begin{equation}
 A_{ij}=\frac{\partial x_{i,retrieved}}{\partial x_{j,true}}.
  \end{equation}
 where $x_{i,retrieved}$ is the retrieved value of the $i^{th}$ parameter and $x_{j,true}$ is the true value of the $j^{th}$ parameter.  The diagonal elements tell us how much a retrieved parameter will respond to an actual change in that parameter in the atmosphere.  For a given change in the true atmospheric state of some parameter $i$, if the measurements are perfect, we would expect to retrieve exactly that same change and hence, the value of $A_{ii}$ would be one.  If the measurements contribute nothing to our knowledge of parameter $i$, that is all of our knowledge of its value is from the prior, then $A_{ii}$ will be zero.  We can use this diagnostic to assess how heavily our error estimations are informed by the measurements.  This is most important when using the Level-by-Level temperature profile retrieval.  The sum of the diagonal elements of this matrix determines the total number of independent pieces of information that can be retrieved from the measurements. This can never be larger than the total number of individual data points.
  
\section{Test on Synthetic Measurements} \label{sec:three}
In this section, we apply the CHIMERA to a set of synthetic measurements in order to assess the robustness of each retrieval algorithm.
\subsection{Synthetic Observations}
\begin{figure*}[h]
\begin{center}
\includegraphics[height=4.5in,width=!, angle=0]{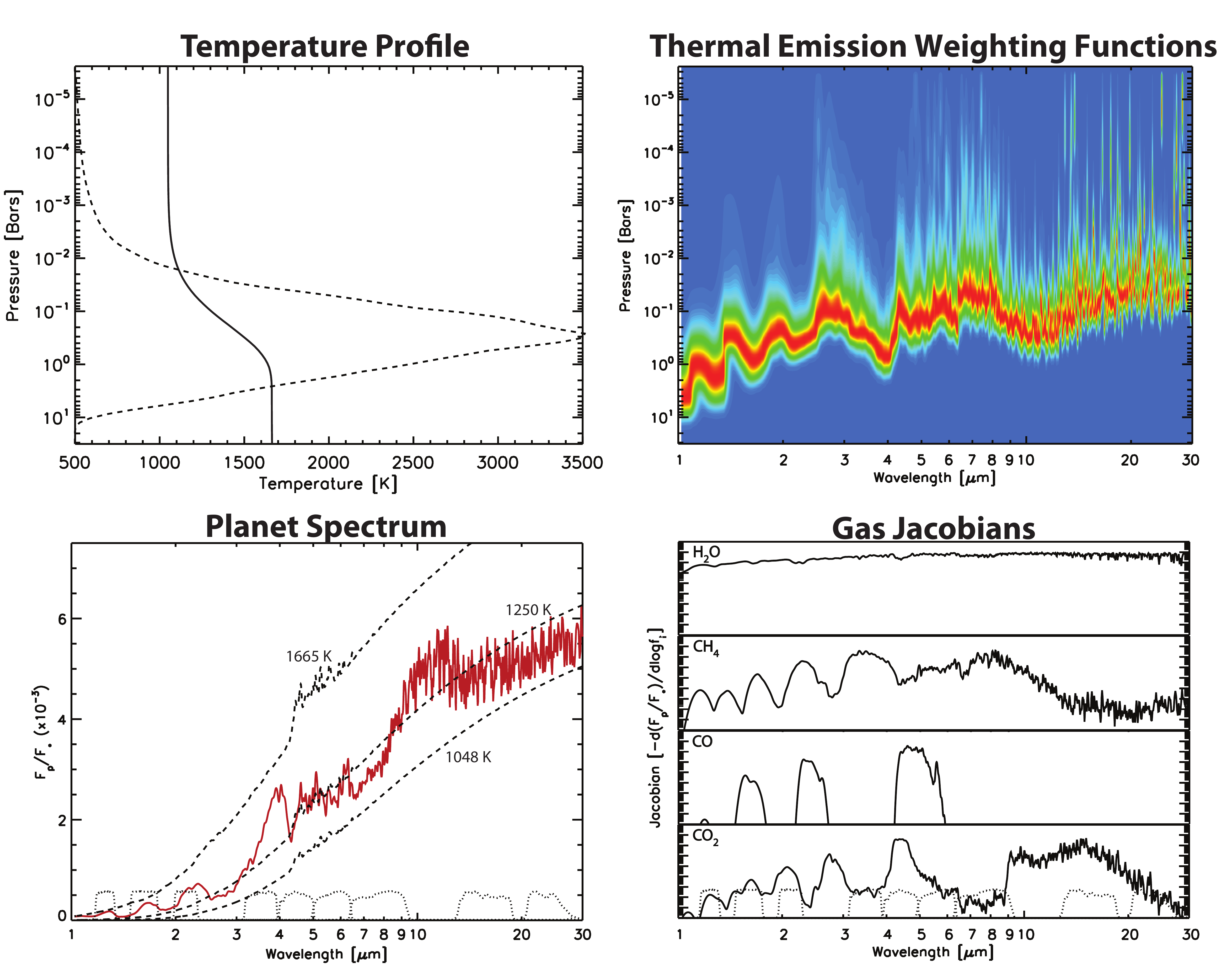}
\end{center}
\caption[Synthetic Planet Atmosphere and Spectrum]{ \label{fig:figure2}Synthetic planet atmosphere and spectrum.  Top-Left: Model temperature-pressure profile.  The solid curve is the temperature profile and the dashed curve is the averaged thermal emission contribution function, or where the emission in the atmosphere is coming from.  The temperature profile is constructed using equations \ref{eq:TP_par}-\ref{eq:tau} and the parameters in Table \ref{tab:table1}.   Top-Right:  Thermal emission contribution function.  This plot shows where the emission is coming from as a function of wavelength, smoothed to a resolution of 0.05 microns.   Red corresponds to the peak of the thermal emission weighting functions, where the optical depth is unity, and blue represents zero emission.  Most emission emanates between a few bars and 0.01 bars with deeper layers probed by shorter wavelengths.  Bottom-Left:  Resulting spectrum smoothed to a resolution of 0.05 microns.   Blackbodies for the hottest, coolest, and average temperatures are shown.  The dotted curves at the bottom are the filter profiles for typical photometric observations.  Bottom-Right:  Gas Jacobian generate from equation \ref{eq:gas_jac}.  This plot shows the sensitivity of the flux contrast as a function of wavelength to the various absorbers (the units are arbitrary but consistent).  }
\end{figure*}

We create a generic hydrogen dominated hot jupiter planet and derive its emission spectrum in three different observing scenarios.  Table \ref{tab:table1} summarizes the basic planet parameters used to generate the model atmosphere and contrast spectrum.  For simplicity we assume that the trace species have mixing ratios that are constant with altitude.  Equations \ref{eq:TP_par}-\ref{eq:tau} are used to generate the atmospheric temperature profile of the planet from the values in Table \ref{tab:table1} .   Figure \ref{fig:figure2} shows the model atmosphere and spectrum of the synthetic planet.  The raw flux is divided by a PHOENIX stellar grid model that closely matches the chosen stellar properties.  The thermal emission contribution functions (Figure \ref{fig:figure2}, top-right) suggest that most of the emission originates between a few bars and a few mbars.  Our synthetic data only provide believable estimates for the temperatures and abundances over this region of the atmosphere.  The thermal contribution functions indicate that the emission from shorter wavelengths comes from deeper layers in the atmosphere and regions of high opacity tend to push the emission to higher altitudes.    In this example, water is the dominant opacity source and acts almost like a continuum absorber across the spectrum (Figure \ref{fig:figure2}, bottom-right ).  If we had no absorbing molecular species other than H$_2$/He most emission would originate from the $\sim$10 bar level.  

We compare the retrieval approaches on only one single fictitious example for illustrative purposes.  Admittedly there are potentially infinite combinations of temperatures and compositions that exist in nature and one example planet does not do that diversity justice.  In a future investigation, currently in progress, Part II, we explore a small set of actual observations of exoplanetary atmospheres that span a wide range of effective temperatures and compositions.    Our conclusions to come regarding the retrieval approaches generally hold on the more diverse set of planetary compositions.

\begin{figure*}[h]
\begin{center}
\includegraphics[height=5.5in,width=!, angle=0]{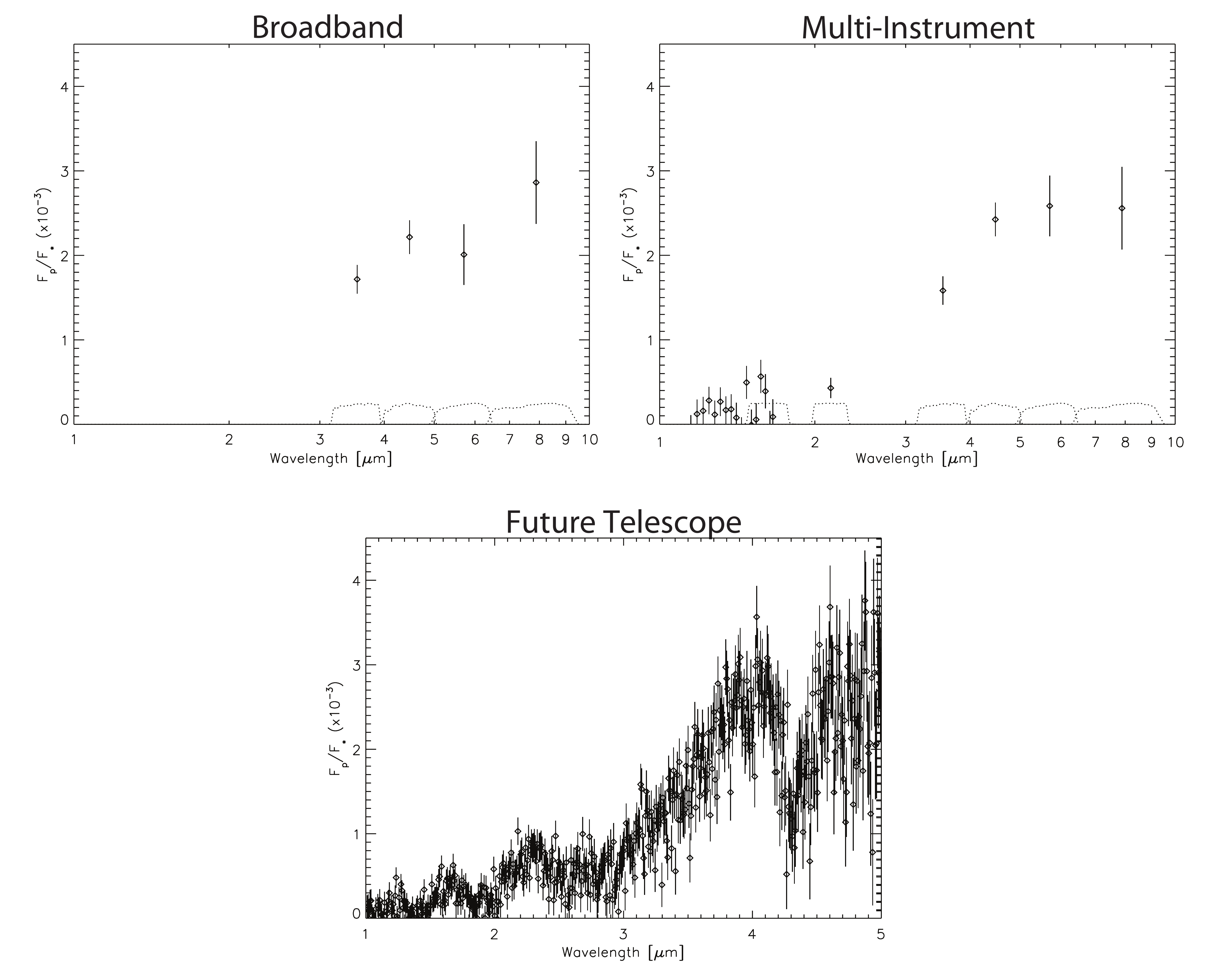}
\end{center}
\caption[Synthetic Observation Scenerios]{ \label{fig:figure3} The spectrum of the synthetic hot jupiter observed in three different scenarios.  These ``observations" are created by convolving the high resolution spectrum in Figure \ref{fig:figure1} with the appropriate instrumental profiles.  Random noise is then added to each data point.  Top-Left:  Synthetic observations as viewed through the Spitzer broadband 3.6, 4.5, 5.7, and 8 $\mu$m channels.  Top-Right: Multi instrument observations that include WFC3 (1.15-1.63 $\mu$m), ground based H and K$_s$, and Spitzer Broadband (3.6, 4.5, 5.7, and 8 $\mu$m). Bottom: Hypothetical future spaceborne observations.  The dotted curves on the bottom of each plot are the photometric transmission functions. }
\end{figure*}

We now create simulated observations for our synthetic planet under three different regimes.   The first regime is a set of broadband observations through four of the Spitzer Infrared Array Camera (IRAC) channels at 3.6, 4.5, 5.7, and 8 $\mu $m (top-left, Figure \ref{fig:figure3}).  This represents the spectral quality that is most commonly available for hot jupiters today.  To create the synthetic observations, the spectrum in Figure \ref{fig:figure1} is first integrated over the IRAC filter functions at each channel and then random noise is added to each channel determined by the error bars size.  The size of the error bars are representative of typical errors of IRAC observations (e.g., Machalek et al. 2009).  

The second observational scenario is a multi-instrument case combining both Spitzer photometry, ground based photometry, and Hubble Wide Field Camera 3 spectra (WFC3) (top-right Figure \ref{fig:figure3}).   This combined set of observations from various instruments is more representative of the current level of observations that can be made today, and likely for the next half-decade, for many planets  (e.g., WASP12b, WASP4b, HD209458b).  Again, we use the same four Spitzer IRAC channels and error bars as before but also include ground based H and K$_s$ band photometry points.  The error bars are taken from Crossfield et al. (2012).  To create the synthetic WFC3 measurements (1.15-1.63 $\mu$m), the high resolution spectrum is convolved with a Gaussian instrumental profile with a FWHM of 0.0325 $\mu$m with error bars taken from Swain et al. (2012).   Random noise is added to each point.  

The third observational scenario illustrates the performance of a potential modest (by modest we mean reasonable cost) future spaceborne, FINESSE-like, telescope (bottom, Figure \ref{fig:figure3}).  These simulated observations are created by convolving the high resolution spectrum with a moderate resolution Gaussian instrumental profile with a FWHM of 0.0075 $\mu$m (R$\sim$300 at 2 $\mu$m).  The measurement error bars, and hence random noise, are only suggestive and are based on a FINESSE-like  noise model (Swain 2012).   This spectral resolution is comparable to that of Exoplanet Characterization Observatory (EchO) below 5 $\mu$m,  but less than the James Webb Space Telescope (JWST) Near Infrared Spectrometer.  

Aside from the potential development of a ground-based near-IR spectroscopy program,  most observations for the foreseeable future are likely to fall somewhere between the first and second cases.     We are also being optimistic in our ``worst" case observational scenario by including four broadband points instead of the now typical two from IRAC.  In the latter case it is impossible to provide any unique constraints on the atmosphere without imposing many pre-conceived assumptions and priors.

\subsection{The Prior}
As mentioned in  $\S$\ref{sec:OE},  the prior is important when the spectral information content is limited.  We use a prior on both the gases and the temperature profile.  For the purposes of this synthetic study, we assume Gaussian priors on the parameters that control the temperature profile and on the gas abundances.    We could have chosen flat (un-informed) priors, however, the formalism of optimal estimation requires that the prior be Gaussian, and hence we maintain this prior for all of the retrieval approaches.  We choose extremely broad Gaussian priors in order to mitigate the influence they have on the retrievals.    For the temperature profile prior, we choose parameters that would reasonably match an atmosphere that is in radiative equilibrium over a wide range of conditions (e.g., variations in $\kappa_{IR}$, $\gamma_{v_{1}}$, $\gamma_{v_{2}}$,  $\beta$, and $\alpha$ ).     Table \ref{tab:table2} lists the prior parameters we use in terms of the prior mean, $\bf x_a$ and the prior covariance matrix, $\bf S_a$.  

In addition to the Gaussian priors, we impose a lower limit on mixing ratios with a value $1\times10^{-12}$ and an upper limit requiring the sum of the mixing ratios of the four retrieved gases to be less than 0.15.   These limits attempt to bound what can be reasonably expected for the compositions of a hydrogen dominated atmosphere.  Also, it would be impossible to detect a gas with an abundance less than 1 ppt in these simulations.   We do not place constraints on the correlations amongst the different molecules.  There are a variety of C to O ratios, metallicities, and disequilibrium effects that can lead to all sorts of combinations of chemical abundances.  Imposing too many constraints would negate the purpose of the retrieval. Nature also has a tendency to surprise us beyond our physical expectations, therefore we feel it would be unwise to impose stringent physical constraints on the correlations amongst the gas abundances.  

We also impose a limit on the parameters that govern the temperature structure.  We do not allow $\kappa_{IR}$ to go above or below 10 and $1\times10^{-4}$, respectively.  The lower limit is roughly the order of magnitude value of the Planck mean opacity expected for an all hydrogen atmosphere.  The upper limit is a bit extreme but would be representative of an extremely opaque atmosphere.  The upper and lower bounds on $\gamma_1$ and $\gamma_2$ are between 10 and $1\times 10^{-4}$ and are chosen to allow for a reasonable span of temperature profiles ranging from ones with inversions to ones nearly transparent to solar radiation.  $\alpha$ can only have physically meaningful values between 0 and 1.  $\beta$ cannot have values below 0, and we impose an upper limit of 2.  A value of unity would be perfect radiative equilibrium with unit emissivity, full redistribution, and zero albedo.  The upper limit would be representative of low redistribution efficiency or a low emissivity.  This value could go higher in the presence of very low emissivity and redistribution.   Generally these upper and lower limits rarely matter as most of the posteriors lie well within their ranges.  

Additionally we do not attempt to self-consistently solve for the opacity parameters and composition for several reasons.  First, we do not know what the visible aborbers are since we are not retrieving them.  Hence, we would not be able to self-consistently solve for $\gamma_1$ and $\gamma_2$ unless we a-priori assumed we knew what those absorbers were, their abundances, and vertical distributions (at which levels do they absorb).  Second, the infrared opacity, $\kappa_{IR}$, is constant with altitude (pressure) in this parameterization.  Generally the opacity may have a pressure dependence, and hence solving self-consistantly for this constant-with-altitude opacity would not actually recover the pressure dependence of this opacity.    Though we tried to choose a physically motivated parameterization, and it is as radiative equilibrium throughout the atmosphere is maintained, the parameters can be thought of simply as free parameters.

\begin{figure*}[h]
\begin{center}
\includegraphics[height=5.5in,width=!, angle=0]{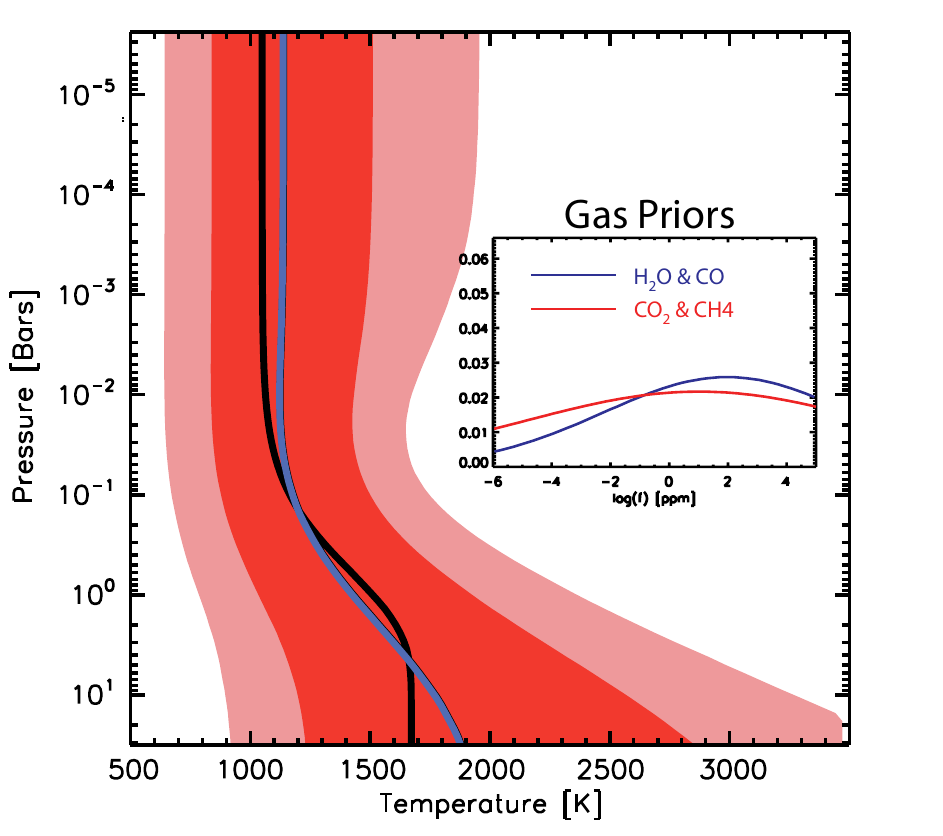}
\end{center}
\caption[Prior]{ \label{fig:figure4} Temperature and gas priors (inset).  Dark red represents the 1$\sigma$ spread in the allowed temperature profiles as a result of the prior parameter distributions in Table \ref{tab:table2}.  Light red is the 2$\sigma$ spread allowed in the temperature profiles.  The blue curve is the median temperature profile and the black curve is the temperature profile constructed from $\bf x_{a}$ in Table \ref{tab:table2}  .  The gas priors are broad Gaussians.  H$_2$O and CO have the same prior mean, CH$_4$ and CO$_2$ have the same prior mean.  Note that the prior is gaussian in {\em log} of the mixing ratios.  The y-axis on the inset is normalized probability. }
\end{figure*}

Figure \ref{fig:figure4} shows the resulting temperature distribution and gas priors (inset).  The prior temperature profile distributions are reconstructed by propagating the Gaussian prior probability distributions (including the above limits) of  $\kappa_{IR}$, $\gamma_{v_{1}}$, $\gamma_{v_{2}}$,  $\beta$, and $\alpha$ in Table \ref{tab:table2} through equations  \ref{eq:TP_par}-\ref{eq:tau}.   Upon reconstructing the temperature profiles there are thousands of temperatures for each pressure level.   With these profiles a histogram of temperatures at each level can be constructed.  Rather than show the ``spaghetti diagram" with thousands of individual profiles, we show the 1$\sigma$ (68$\%$) and 2$\sigma$ (95 $\%$) confidence intervals at each pressure level.   These confidence intervals are what is shown in Figure \ref{fig:figure4}.

\subsection{Results from the Parameterized Temperature Profile}

\begin{figure*}[h]
\begin{center}
\includegraphics[height=5.5in,width=!, angle=0]{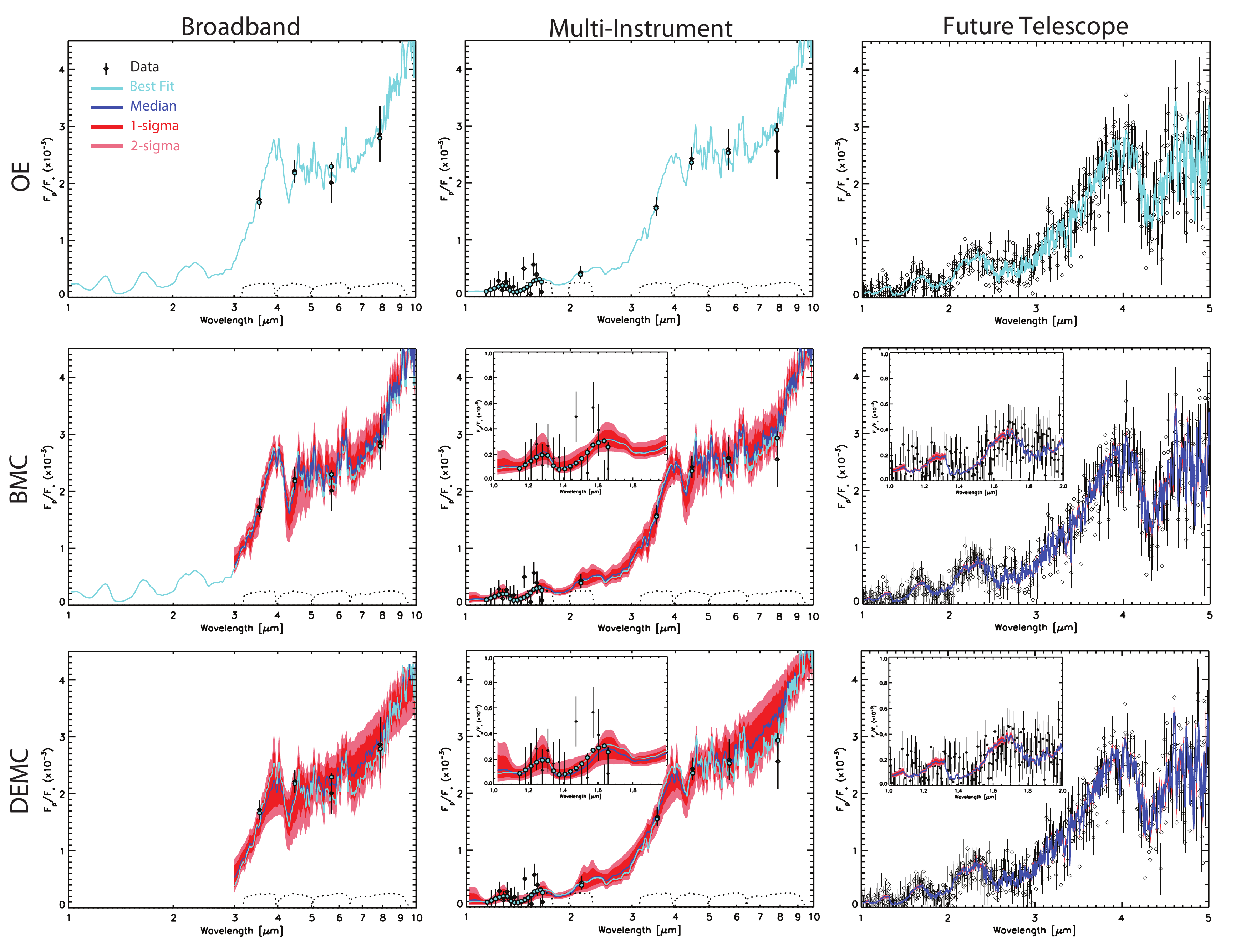}
\end{center}
\caption[Best Fit Spectra]{ \label{fig:figure5}  Fits to the three different sets of data (columns) from each of the three different retrieval techniques (rows). The first scenario consists of the four IRAC photometry channels.  The second scenario consists of the four IRAC photometry channels, ground based H and K$_s$ band photometry, and HST WFC3 spectroscopy.  The third scenario is representative of a FINESSE-like future, spaceborne telescope.   The best fits from each scenario and technique are shown in light blue.  The light-blue circles with the black borders are the best fits binned to the data.     The chi-squared per datapoint from the optimal estimation best fit broadband scenario, multi-instrument scenario, and future telescope scenario are respectively, 0.197, 0.686, and 0.955.   The bootstrap Monte Carlo and the differential evolution Markov chain Monte Carlo approaches generate many thousands of spectra.  The median of these spectra is shown in blue and the 1- and 2$\sigma$ spread in the spectra are shown in dark and light-red, respectively.  The best fit from the thousands of spectra are shown in light blue.  The best fit reduce-chi-squares from BMC and DEMC are of similar values to those from OE.   The dotted curves at the bottom of each panel are the broadband filter transmission functions.  The insets are a zoom in of a spectral region between 1and 2 $\mu$m to better show the spread in the spectra.   Note that there is virtually no spread in the spectra for the future telescope case.        }
\end{figure*}

We apply the three retrieval techniques to the three synthetic observations in Figure \ref{fig:figure3} under the radiative equilibrium temperature profile parameterization.  This temperature profile has few parameters (5) because of the assumption of radiative equilibrium.   The temperature prior in Figure \ref{fig:figure4} is used in all three techniques for all three retrieval cases.     In each case, we retrieve the parameter distributions for 
the following state vector:
\begin{eqnarray}
{\bf x}=[log(f_{H_2O}), log(f_{CH_4}), log(f_{CO}), log(f_{CO_2}), \nonumber\\
  log(\kappa_{IR}), log(\gamma_{v_{1}}), log(\gamma_{v_{2}}),  \beta, \alpha  ]^T.
\end{eqnarray}
where again, the $f_i$s are the altitude independent volume-mixing ratios.  We start by first applying the optimal estimation approach.  In order to ensure that the retrieval does not get stuck in a local minimum, multiple starting guesses are used.  These typically all converge to the same temperature and gas solution.  As described in $\S$\ref{sec:DEMC}, the covariance matrix and the best fit from OE are then used to initialize the DEMC chains.  Finally, the best fit from OE is used to initialize the synthetic measurement realizations used in the BMC.  Figures \ref{fig:figure5}-\ref{fig:figure8} and Table \ref{tab:table3} summarize the retrieval results and form the basis for the comparisons.  The bounds quoted in Table \ref{tab:table3} are for the 68$\%$ confidence intervals.  

We must be careful in interpreting the confidence interval values when the posteriors extend to the imposed upper and lower limits, especially when those limits are somewhat arbitrary. Parameters with posteriors that approach the imposed lower limit will result in an overestimate of the lower bound on the confidence interval and an underestimate in the upper bound due to the imposed upper limit.  In some cases if there were no Gaussian prior or no lower limit, the lower bound could extend to $-\infty$!  Of course we would interpret such a case as only having an upper bound.  

 Figure \ref{fig:figure5} shows the spectral fits.  The first row shows the single best fit from optimal estimation.    The second and third rows show the fits from the BMC and DEMC, respectively.  Since the BMC and the DEMC provide many thousands of spectra, rather than plot each one, we summarize the fits by showing the median spectrum along with the 1- and 2$\sigma$ spread at each wavelength.  Illustrating the fits in this manner is more representative of the posterior than plotting spectra of different chi-squared levels.  In other words, if a random set of parameters is drawn from the posterior, there would be a $95\%$ chance that the flux at any one wavelength of the spectrum resulting from that parameter draw would fall with in the 2$\sigma$ spread etc.   We also show the best fits as determined by BMC and DEMC.  These best fits, while different than the best fit from OE, are of negligible difference both in terms of the best fit state vector and cost function value.  There is little if any spread in the fits as the measurement quality improves.    In the following subsections we summarize posteriors for the gas compositions, temperatures, and C to O ratios for each observational scenario.

 \subsubsection{Gas Abundance Retrievals} \label{sec:gas_retrieval_par}
\begin{figure*}[h]
\begin{center}
\includegraphics[height=5.5in,width=!, angle=0]{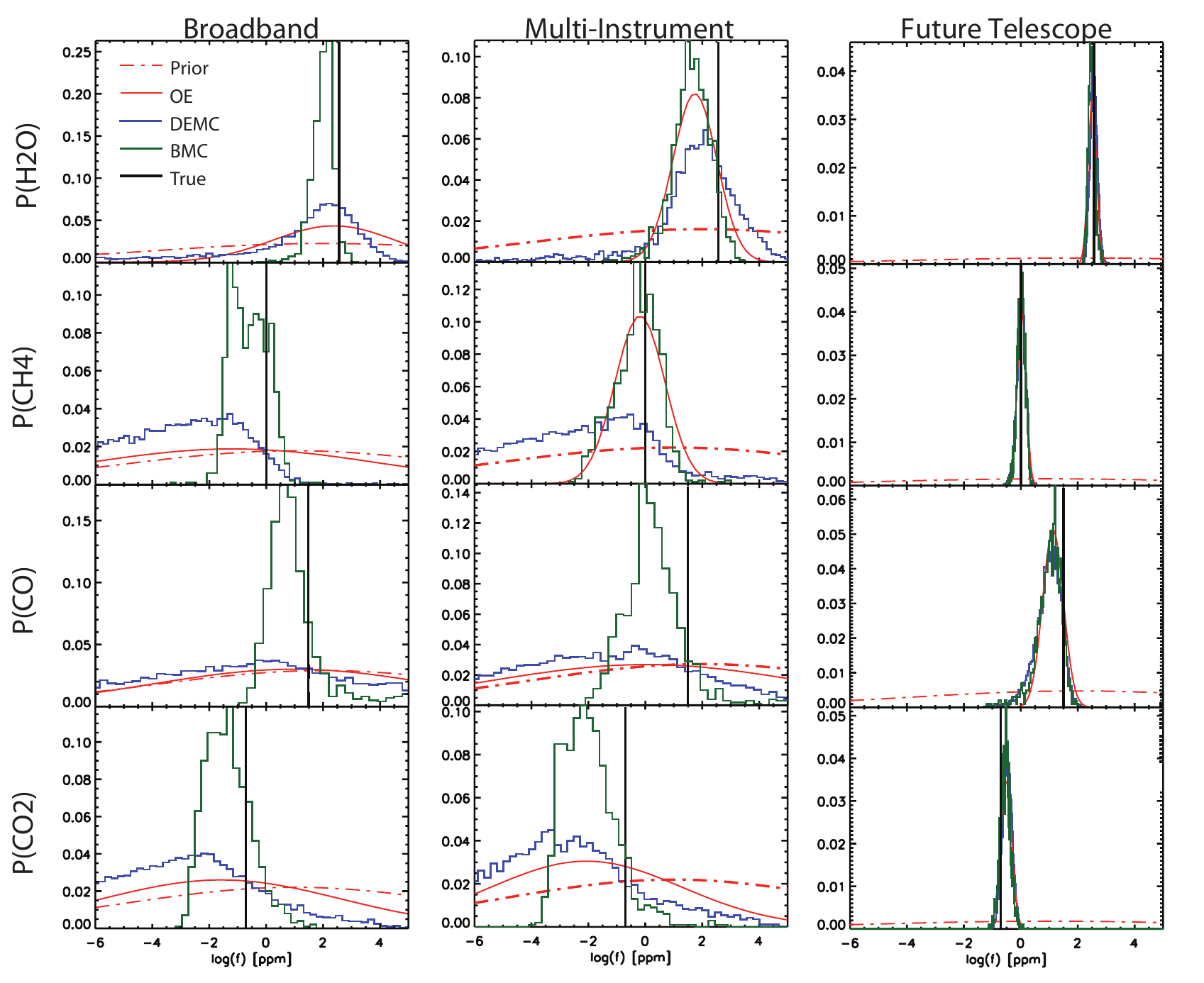}
\end{center}
\caption[Abundance Error Distributions]{ \label{fig:figure6} Marginalized posterior probability distributions for each of the retrieved gases (rows) and observational scenario (columns).   In each panel the probability distribution for each retrieval technique are shown in different colors.  The Gaussian probability distributions from optimal estimation are in red, differential evolution Markov chain Monte Carlo in blue, and bootstrap Monte Carlo in green.  The priors for each gas are the dot dashed red curve.  The true answer is the vertical black line.    }
\end{figure*}

\begin{figure*}[h]
\begin{center}
\includegraphics[height=6.in,width=!, angle=0]{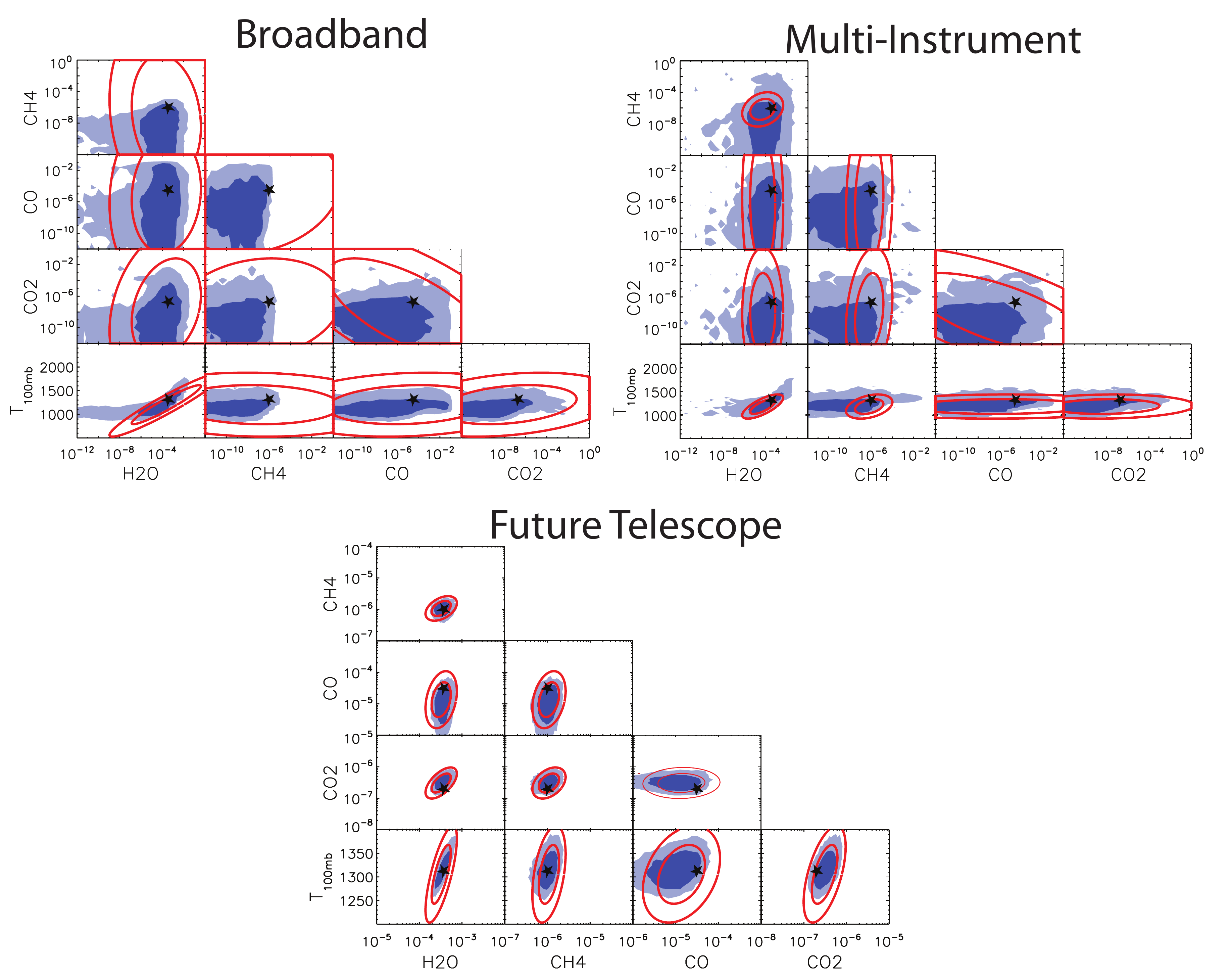}
\end{center}
\caption[Abundance Correlations/Degenracies]{ \label{fig:figure7} Gas correlations for each of the observing scenarios.    The red curves in each are the analytic confidence intervals from the optimal estimation posterior covariance matrix ($\bf \hat S$).  The inner ellipses are the 1$\sigma$ (68$\%$) and the outer ellipses are the 2$\sigma$ (95$\%$) confidence interval.  The 1- and 2$\sigma$ confidence intervals derived from the differential evolution Markov chain Monte Carlo are shown in dark and light blue, respectively.  Note that the scales for the confidence intervals derived from the broadband observations (top) and the multi-instrument observations (middle) are the same.  The scale on the future telescope (bottom) confidence intervals is much smaller.   }
\end{figure*}

\begin{figure*}[h]
\begin{center}
\includegraphics[height=7.5in,width=!, angle=0]{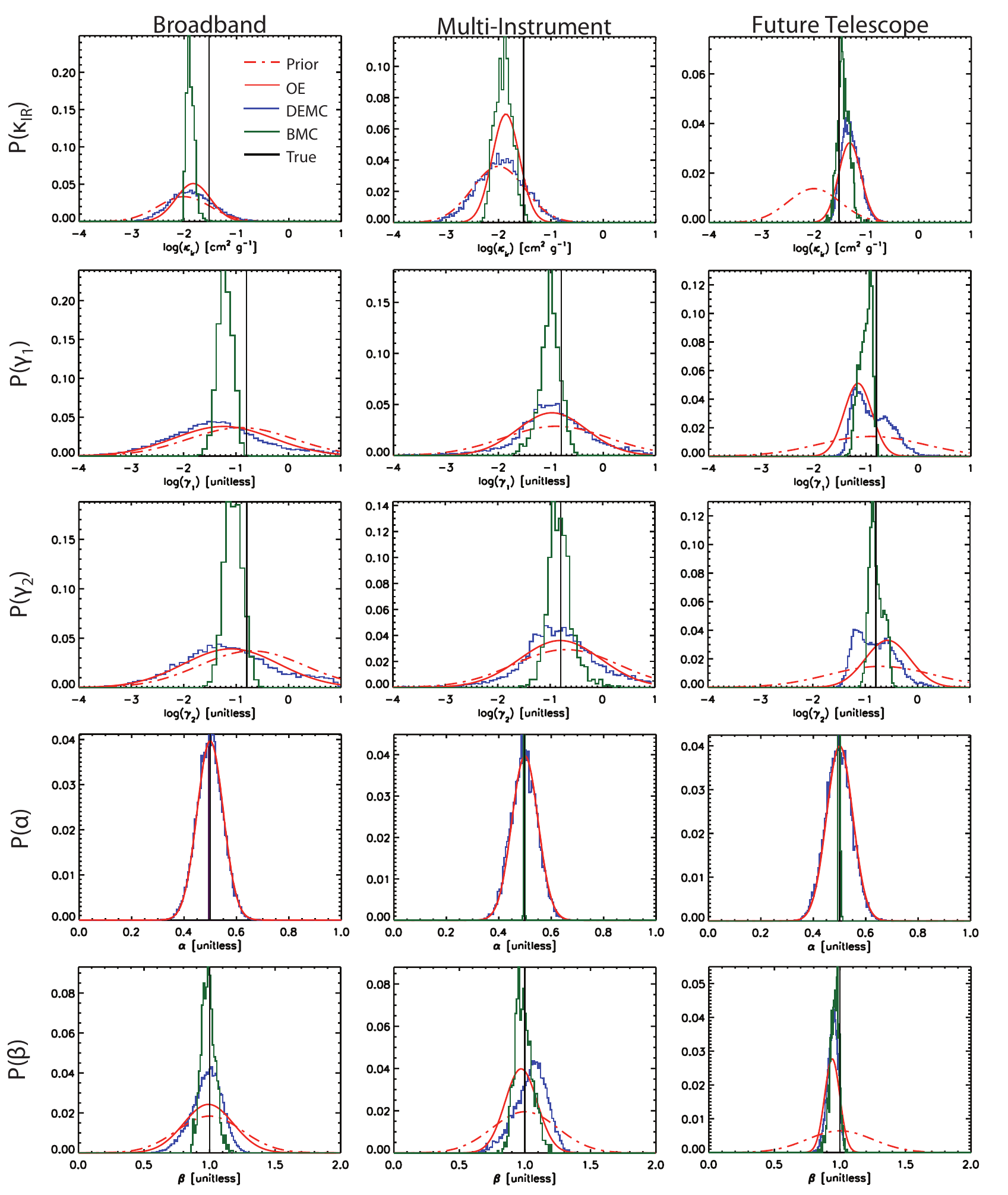}
\end{center}
\caption[Temperature Parameter Error Distributions]{ \label{fig:figure6b} Marginalized posterior probability distributions for each of the retrieved temperature parameters (rows) and observational scenario (columns).   In each panel the probability distribution for each retrieval technique are shown in different colors.  The Gaussian probability distributions from optimal estimation are in red, differential evolution Markov chain Monte Carlo in blue, and bootstrap Monte Carlo in green.  The priors for each gas are the dot dashed red curve.  The true answer is the vertical black line.    }
\end{figure*}

 The gas mixing ratio retrieval results are summarized in Figures \ref{fig:figure6} and  \ref{fig:figure7}.  Figure \ref{fig:figure6} shows histograms of the marginalized posterior for each of the four gases as a result of each retrieval approach and observational scenario.  We take the DEMC posterior (blue) to be representative of the true posterior.  The optimal estimation posteriors (solid red curve) and the prior (dot-dashed red curve) are smooth because they are constructed analytically from the diagonal elements of $\bf \hat S$.   Figure \ref{fig:figure7} is a ``stair-step" plot that shows the correlations amongst the four gases and the gases with temperature comparing optimal estimation to the DEMC.  For brevity, we do not show the bootstrap Monte Carlo correlations.  The solid blue filled regions are the 1 (dark) and 2 (light) $\sigma$ confidence intervals derived from the DEMC, and the red curves are the 1$\sigma$ (inner) and 2$\sigma$(outer) confidence intervals derived from OE. 
  
 The first column of Figure \ref{fig:figure6} shows the marginalized gas posteriors for the broadband observational scenario and the top set of panels of Figure \ref{fig:figure7} show the correlations amongst the gases.   In this observational scenario, the three retrieval techniques produce quite different posteriors.   H$_2$O has a fairly narrow posterior (relative to the prior) near the true state, suggesting that it is reasonably well constrained, at the 1$\sigma$ level, by even this low information content spectra.  This is unsurprising as water is prevalent across all of the channels in this particular spectrum (see Jacobian in Figure \ref{fig:figure2}).   At the 2$\sigma$ level, however, the OE retrievals provide less of a constraint (Figure \ref{fig:figure7}).   The CH$_4$ and CO$_2$ posteriors abruptly fall off towards their upper end suggesting an upper bound constraint on these gases.  The low end of their posteriors begin to track the prior down to the imposed lower limit indicating that from this observational scenario, there really is no observable lower limit to the abundances of these species.  CH$_4$ has a better defined upper edge than CO$_2$ because both the 3.6 $\mu$m and 8 $\mu$m channels overlap with the strongest methane absorption bands.  CO is virtually unconstrained by the synthetic broadband measurements as it closely matches the prior across the full range of values.  The difficulty in constraining CO and CO$_2$ is due to the inability of the 4.5 $\mu$m broadband photometric measurement to decouple the CO and CO$_2$ strong bands, at least for this particular combination of compositions.  It is possible that either of these species may be better well constrained if their abundances are higher.  There is a slight hint of an inverse correlation, as expected, from the optimal estimation results in Figure \ref{fig:figure7}, however this correlation is not apparent from the DEMC results.    
 
Our implementation of optimal estimation struggles to appropriately capture the errors in this observational scenario.   This is because it approximates the posteriors with broad Gaussians which simply do not capture the appropriate structure.  It does, however, do a fairly good job of determining the true state.   The Gaussian approximation cannot appropriately handle upper bounds on CH$_4$ and CO$_2$, causing an overestimate of the 1$\sigma$ upper bound.  We note, that at least it is overestimating the errors rather than underestimating them.  OE does a fine job at approximating the posterior for CO, which happens to be similar to the prior, again suggesting no constraint.  This is reaffirmed by looking at the averaging kernel element, $A$, in Table \ref{tab:table3} which shows that most (~70$\%$) of the information in constraining CO comes from the prior (see \S \ref{sec:TP_approach}).    The bootstrap Monte Carlo (green curve) completely fails to appropriately capture the posterior in this particular observational scenario. This is because the different realizations from the BMC approach produce parameter distributions that are confined to a small area of phase space localized near the best fit solution from optimal estimation and cannot therefore sample the entirety of the posterior.  
 
The second column of Figure \ref{fig:figure6} shows the gas posteriors for the multi-instrument observational scenario.  The information gain from this synthetic observational scenario is only marginal relative to the broadband case.  Water has the largest improvement in uncertainty due to the leverage provided by the WFC3 spectra which covers the 1.15 $\mu$ and 1.4 $\mu$m water bands.   Upon inspecting Figure \ref{fig:figure7} we find that the WFC3 data combined with the ground based photometric points trims the 2$\sigma$ tail but does little to improve the 1$\sigma$ uncertainties.  Sadly, there is virtually no reduction in the uncertainties on CH$_4$, CO, and CO$_2$.  In fact the marginalized posteriors (Figure \ref{fig:figure6}) produced by DEMC in this observational scenario look nearly identical to the previous case.   Optimal estimation provides an accurate error estimation for water but  appears to provide an overly optimistic estimation of the uncertainties on CH$_4$.  OE, as in the previous scenario, captures the essence of large uncertainties on CO and CO$_2$ with broad Gaussians. bootstrap Monte Carlo underestimates the uncertainties in all species with the exception of water.  

Finally, results for a hypothetical future spaceborne telescope are shown in the last column of Figure \ref{fig:figure6} and bottom set of panels of Figure \ref{fig:figure7}.  The reduction in uncertainties are staggering when compared to the previous observational scenarios.  All of the gases are constrained to within better than an order of magnitude.  This is an orders-of-magnitude improvement over the previous cases.   The high signal-to-noise and high spectral resolution combine to provide excellent coverage of each of the four gases.  With this high quality spectrum, all three retrieval approaches give the same results.  The Gaussian assumption used in optimal estimation is perfectly appropriate in this case. The differences in the 1$\sigma$ uncertainties derived from OE are less than $\sim10\%$ than the uncertainties derived from DEMC.  The two-dimensional confidence intervals in Figure \ref{fig:figure7} also agree quite well.  The prior also plays very little role in the retrieval as shown with the near unit averaging kernel elements in Table \ref{tab:table3}.

  \subsubsection{Temperature Profile Retrievals}\label{sec:TP_retrieval_par}
The marginalized posteriors for the five parameters that govern the shape of the temperature profile for each observational scenario are shown in Figure \ref{fig:figure6b}.  We find very little sensitivity in all scenarios, e.g., the retrieved posterior is just the prior, to the $\alpha$ parameter that governs the partitioning between the two downwelling visible streams.  We suspect this would not always be the case especially if a thermal inversion exists.  In the broadband scenario it appears that the first four temperature parameters are un-informed by the data.  Again, the posterior is the prior (with the BMC retrieval technique severely underestimating the uncertainties on these parameters relative to DEMC and OE).  There is more sensitivity to the $\beta$ parameter since that controls the overall magnitude of the temperature profile.  At higher spectral resolution the agreement amongst the techniques is not quite as good as it is in Figure \ref{fig:figure6}.  We owe this to the strong nonlinearities in the temperature profile parameterization.   There are a variety of combinations and distributions of parameters that can recreate the same spread in temperatures over the pressure levels of which the spectrum is sensitive.  Because of this, we believe it is more constructive and physically meaningful to look at the resultant temperature profiles themselves.  

\begin{figure*}[h]
\begin{center}
\includegraphics[height=6.5in,width=!, angle=0]{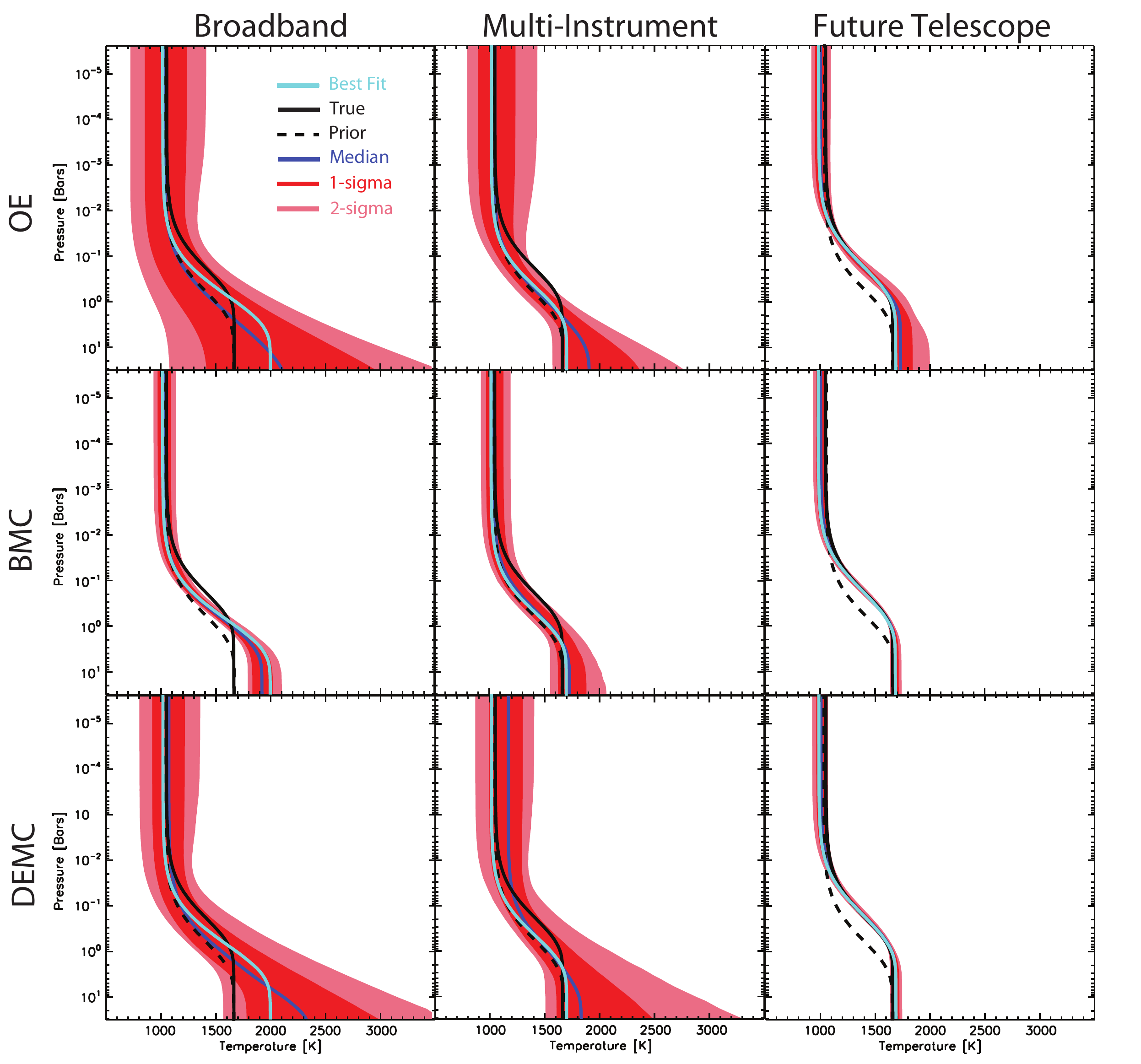}
\end{center}
\caption[Temperature Profile Uncertainties]{ \label{fig:figure8} Temperature profile posteriors for each observational scenario (columns) and each retrieval technique (rows).   The solid black curve in each panel is the true temperature profile constructed with equations \ref{eq:TP_par}-\ref{eq:tau} and the parameters in Table \ref{tab:table1}.  The dashed black curve is constructed from the temperature parameters, $x_a$, just as in Figure \ref{fig:figure4}   The blue curve is the median temperature profile.  The light blue curve in each panel is the best fit temperature profile for the corresponding scenario and technique.  The dark and light red regions are the 1- and 2$\sigma$ ($68\%$ and $95\%$) uncertainties, respectively.         }
\end{figure*}

We can reconstruct the temperature profiles by randomly drawing state vector parameters from the posterior, derived from any of the techniques. Figure \ref{fig:figure8} shows range of temperature profiles that can be reconstructed.  The dark and light red swaths show the 1 and 2$\sigma$ bounds on the reconstructed temperature profiles.   The blue curve is the statistical median of these profiles and the light blue curve is the best-fit temperature profile from each scenerio.  This median profile not representative of any one given temperature profile, and in fact this median profile may not even provide a good fit to the observations or adhere to the parameterization,  but is shown simply as a statistical summary of all possible temperature profiles.  The black curve is the true temperature profile.

The temperature profile posteriors for the broadband scenario (first column Figure \ref{fig:figure8}) from OE and DEMC have similar widths and both capture the entire true temperature profile within the 2$\sigma$ interval.    There is also a non-negligible ($\sim$30-50$\%$) reduction in the temperature precision compared with the prior, over the atmospheric region probed by these observations.  Outside of the range spanned by the thermal emission contribution functions (Figure \ref{fig:figure2}), the temperature uncertainty grows and begins to relax back towards the prior where there is no observational constraint.  Again, the BMC approach completely underestimates the error when compared with the other two approaches because of its inability to fully characterize the posterior outside of a small region of phase space localized around the OE original best fit.  

Moving onto the multi-instrument observations (middle column Figure \ref{fig:figure8}) we find a 22\%  reduction in the temperature uncertainty between 1 and 0.01 bars. The large number of spectral channels from the WFC3 data that have weighting functions over this region are the primary contributors to this increased precision.  OE and BMC underestimate the temperature uncertainties relative to DEMC at 100 mbars, but the OE and DEMC have reasonable agreement over the entire profile.  

The future spaceborne telescope observations improve the temperature uncertainties by a remarkable factor of $\sim4.5$ over the previous case.  optimal estimation slightly overestimates the temperatures outside the atmospheric levels probed by the observations.   This overestimate is due to the overestimation of the $\gamma_{v1}$ and  $\gamma_{v2}$ posterior widths.  These two parameters control the relative difference between the upper atmosphere and lower atmosphere temperatures.  Hence, extreme values of $\gamma_{v1}$ and $\gamma_{v2}$ will affect these regions more than the middle atmosphere.  This why the OE and the DEMC temperatures agree in the middle atmosphere but not outside of it. 

We also show the correlations of the molecular abundances with temperature in Figure \ref{fig:figure7}.  Rather than show the correlations of the molecular abundances with the five temperature profile parameters, we choose to show how they correlate with a more physically useful quantity, the 100 mbar temperature, which again, where most of the thermal emission emanates.  In this particular scenario, the spectrum is dominated by water absorption, as is the case for most hot-jupiters with solar-like composition.  This large abundance of water acts almost like a continuum absorber, e.g., it absorbs everywhere in our synthetic observations.   This results in a strong correlation with temperature.  If the water abundance increases the temperature must also increase to maintain the same upwelling flux, and vice-versa.   This correlation is prevalent in all three observational scenarios, as even the broadband points are strongly effected by water vapor absorption (see Figure \ref{fig:figure2}).  However, if water is at a lower abundance, as can bee seen in the 2$\sigma$ tail in the broadband scenario, the correlation disappears because the strength of the continuum-like absorption becomes less prevalent.  In the two lower resolution observational scenarios the other molecular abundances have less of a correlation since their absorption is less prevalent.  However, as the spectral resolution increases the absorption bands of these other molecules are better resolved allowing for a stronger correlation with temperature.  In in cooler (T $<$ 1000 K) planets methane will be more prevalent.  When methane is abundant it has a similar continuum-like impact on the spectrum (see Figure \ref{fig:figure1}).  This will result in a strong correlation with temperature just as it is with water, even in the lower resolution scenarios.   We can see hints of the happening towards the upper limit of the methane abundance in broadband scenario in Figure \ref{fig:figure7}.

 \subsubsection{C to O Ratios}\label{sec:CtoO}
\begin{figure*}[h]
\begin{center}
\includegraphics[height=4.5in,width=!, angle=0]{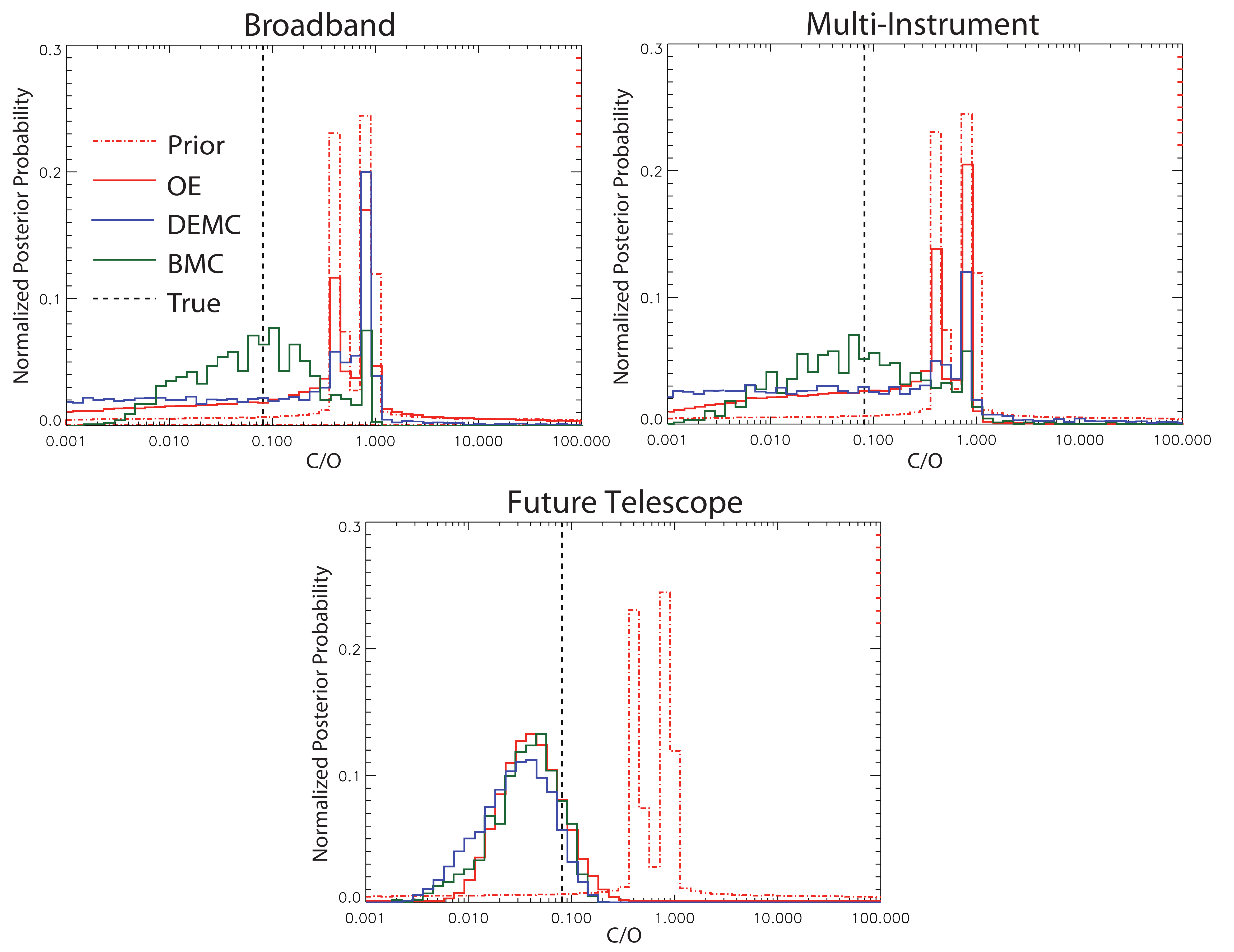}
\end{center}
\caption[C/O Ratio Uncertainties]{ \label{fig:figure9} C to O ratio posteriors.  The dot-dashed red curve is the prior, the solid red curve is from OE, blue is from DEMC, and green is from BMC.  The vertical dashed line is the true C/O.  The top left panel are the C/O's derived from the broadband observational scenario, top right are the C/O's derived from the multi-instrument scenario, and the bottom are the C/O's derived from the future spaceborne telescope scenario. Though it appears that the BMC characterizes the C/O errors well, it is for the wrong reasons.  See  \S \ref{sec:CtoO}      }
\end{figure*}

Determination of the C to O ratios of explanatory atmospheres is critical to the understanding of their atmospheric chemistry (Lodders \& Fegley 2002;  Moses et al. 2012) and formation environments ({\"O}berg, Murray-Clay, \& Bergin 2011; Madhusudhan et al. 2011).   Given the abundance posteriors derived with CHIMERA, we can compute C/O posterior distributions (Figure \ref{fig:figure9}) .  The C/O is calculated with the following formula,
\begin{equation}\label{eq:CtoO}
C/O=\frac{\Sigma C}{\Sigma O}\approx \frac{CH_4+CO+CO_2}{H_2O+CO+2CO_2}.
\end{equation}
There are a few simple things to note about this equation.  When CO is the dominant species, C/O  is 1.  If CO$_2$ is the dominant species, C/O will be $1/2$.  When methane dominates the C/O will be large and when H$_2$O dominates C/O will be small.  The solar C/O is ~$0.55$.  A number of exoplanets have reported C/O's near 1.  We can construct the C/O probability distributions by propagating the probability distributions of each gas through equation \ref{eq:CtoO}, similar to the method used to construct the temperature profile posteriors.  Before inspecting the posteriors derived from CHIMERA, we find it illustrative to investigate the prior.   Upon propagating the Gaussian priors (with the limits) of the gases through equation \ref{eq:CtoO} we obtain the C/O prior in Figure \ref{fig:figure9}.  We find that this prior has two peaks,  one is at a C/O of 1 and the other is at a C/O of 0.5.  The locations of these peaks are insensitive to whether or not the gas abundance priors are uniform or broad Gaussians.  These peaks are also insensitive to the lower and upper bounds placed on either a uniform or Gaussian gas prior.  These double peaks are due to an elegant mathematical misfortune.  For illustrative purposes, let us assume we draw the set of four gases from a uniform log prior.  We would expect then, that one gas will have a larger abundance than the other three 25$\%$ of the time.  This means that in roughly 25$\%$ of the draws CO will dominate, which would cause the ratio in equation \ref{eq:CtoO} to be one, 25$\%$ of the time.  A similar argument goes for CO$_2$ resulting in a C/O of 0.5 roughly 25$\%$ of the time.  So we see, that if we did not observe a particular planet, and assumed uniformed priors on each of the gases, we would naturally conclude that the planet has equal chances of having a C/O of one or one-half.   However, our priors are not uniform, but rather broad Gaussians, but they are broad enough that this behavior still occurs, with a slight preference for a C/O of one.  

If we observe this planet with the four broadband points we obtain the posteriors in the upper-left panel of Figure \ref{fig:figure9}. OE and DEMC produce similar C/O posteriors both of which maintain the two peaked features at 1 and 0.5 but with overall less power in the peaks.  These features persist simply because the gas posteriors from DEMC and OE do not deviate too strongly from the prior.  There is more probability in the lower C/O tail than in the prior because of the higher values of H$_2$O preferred by the measurements over the prior.  This is good, since the C/O for our fictitious planet is much less than one or one-half.    BMC anomalously captures the true C/O at the peak of its posterior.  Again, BMC greatly underestimates the posterior widths because it only searches a localized region about the OE best fit.  Since the OE best fit gas abundances are very near truth (Figure \ref{fig:figure6}), the BMC posteriors, which are highly localized about the OE best fit parameters, will overemphasize the C/O derived by that best fit.   

The story is the same for the multi-instrument observational scenario.  Unfortunately, at least in this example, it appears that the WFC3 and ground based data provide very little additional constraints in reducing the C/O uncertainty, with the double peaked feature from the prior persisting in the DEMC results.

Improving the observational quality further with a future spaceborne telescope essentially obtains the correct value to high precision.   The peak of the posterior is far enough away from the double peaked prior that the results appear to be less influenced by the prior than the previous cases.  All three retrieval approaches give a nearly identical posterior.

From this exercise we have learned that it is difficult to constrain the C to O ratio of an exoplanet atmosphere.  Simple, uninformed, or nearly uninformed priors on the gas abundances produce a double peaked C/O prior at near solar value and one.  Even in the best cases, current observations are likely to provide only an upper limit on this quantity.  This result suggests that previously published claims to detect enhanced C to O ratios with photometry alone may be influenced by these subtle biases, and should be viewed with strong skepticism.  We will discuss this issue in the context of specific exoplanets in a companion paper.

\subsection{Results from the Level-by-Level Temperature Profile}\label{sec:Level-by-Level}
The Level-by-Level temperature profile approach attempts to determine the temperature for each model layer.   In contrast to the parameterized temperature profile which had only 5 parameters, the Level-by-Level approach requires as many parameters as model layers, for a total of 90 parameters.  This larger number of parameters is far greater than the number of meaningful constraints provided by most observations.  However, this approach makes no assumptions about the physical structure of the temperature profile (e.g., radiative, radiative-convective, advection etc.).  While there is no potentially biasing parameterization, the retrievals can result in unphysical temperature profiles.   Obviously, the temperature at each of the 90 levels cannot be perfectly retrieved, but rather the retrievals have to depend on the prior when spectral information on the temperature is sparse.  

For the Level-by-Level prior we assume an a prior variance of 400 K and covariance amongst each level with all other levels given by equation \ref{eq:TP_covar}.   The 400 K variance is used to produce a similar temperature profile prior as in Figure \ref{fig:figure4}. This correlation helps reduce the effective number of levels that have to be independently retrieved.  Admittedly, the degree of correlation is somewhat of an external arbitrary parameter, but it is chosen to avoid over-fitting (i.e., fitting to the noise) without hindering the Level-by-Level flexibility.  It can be thought of as a smoothing, or more specifically, a regularization.  We can also use the averaging kernel profile to assess where the temperature is constrained by the measurements versus the prior.     The gas priors are the same as before.  We choose only to compare the results from optimal estimation and the bootstrap Monte Carlo.  We do not attempt the DEMC approach on such a large ($\sim$100) number of parameters, as MCMC algorithms (to the best of our knowledge) are not well suited for large numbers of parameters because of the large number of steps required to fully map the $n$-dimensional probability distribution when $n$ is large.

\begin{figure*}[h]
\begin{center}
\includegraphics[height=3.75in,width=!, angle=0]{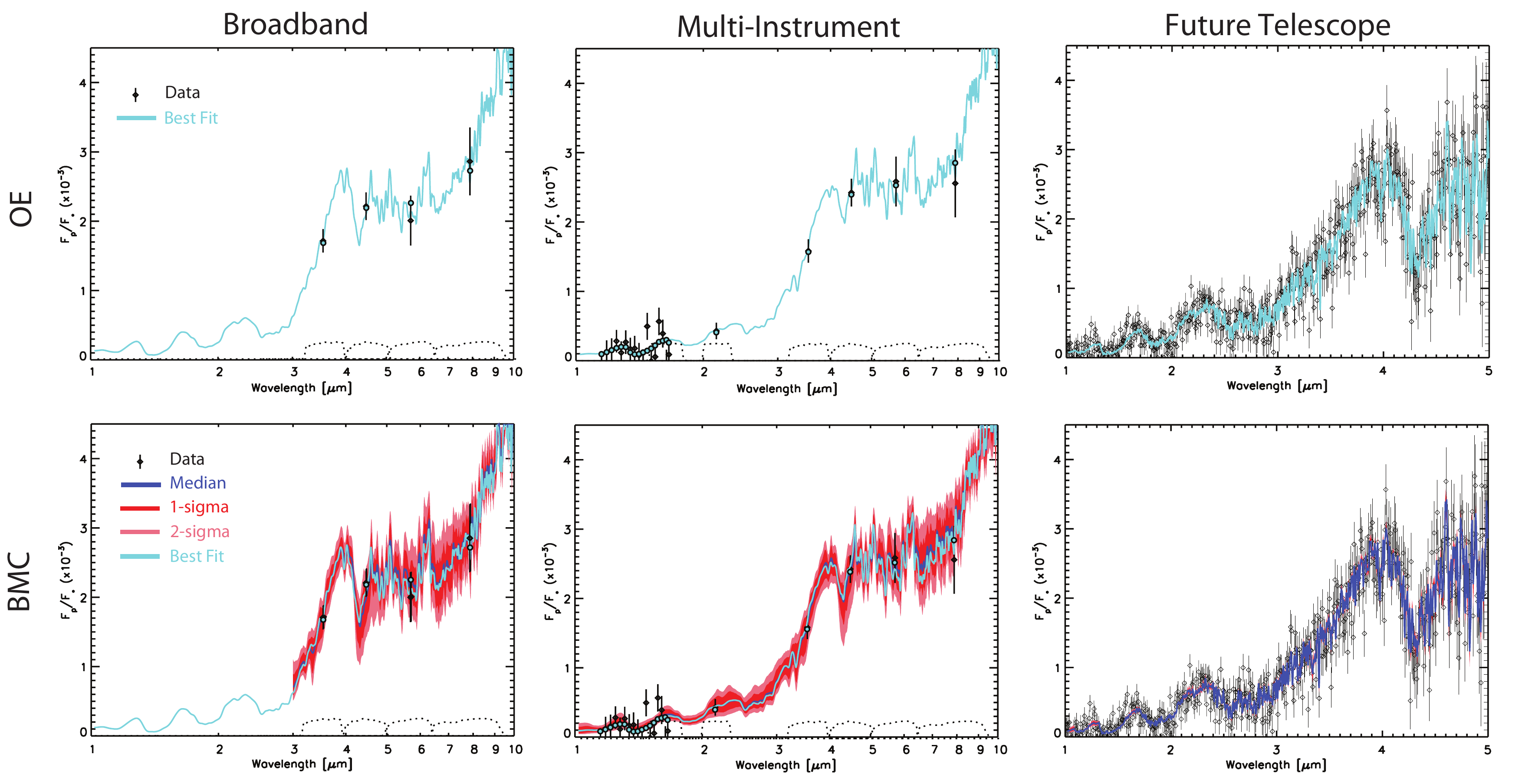}
\end{center}
\caption[Level-by-Level Retrieval Best Fit Spectra]{ \label{fig:figure10}   Fits to the three different sets of data (columns) from two of the retrieval techniques (rows) using the Level-by-Level temperature approach.  The best fits are shown in light blue.  The optimal estimation best fit in each observational scenario is the global best fit.  The light-blue circles with a black border are the best fit spectra binned to the data.  The chi-squared per data point values for the broadband, multi-instrument, and future telescope scenarios are respectively, 0.155,  0.665 , and 0.948.   The bootstrap Monte Carlo approach generates many thousands of spectra.  The median of these spectra is shown in blue and the 1- and 2$\sigma$ spread in the spectra are shown in dark and light-red, respectively.  The dotted curves at the bottom of each panel are the broadband filter transmission functions.      }
\end{figure*}

\begin{figure*}[h]
\begin{center}
\includegraphics[height=5.5in,width=!, angle=0]{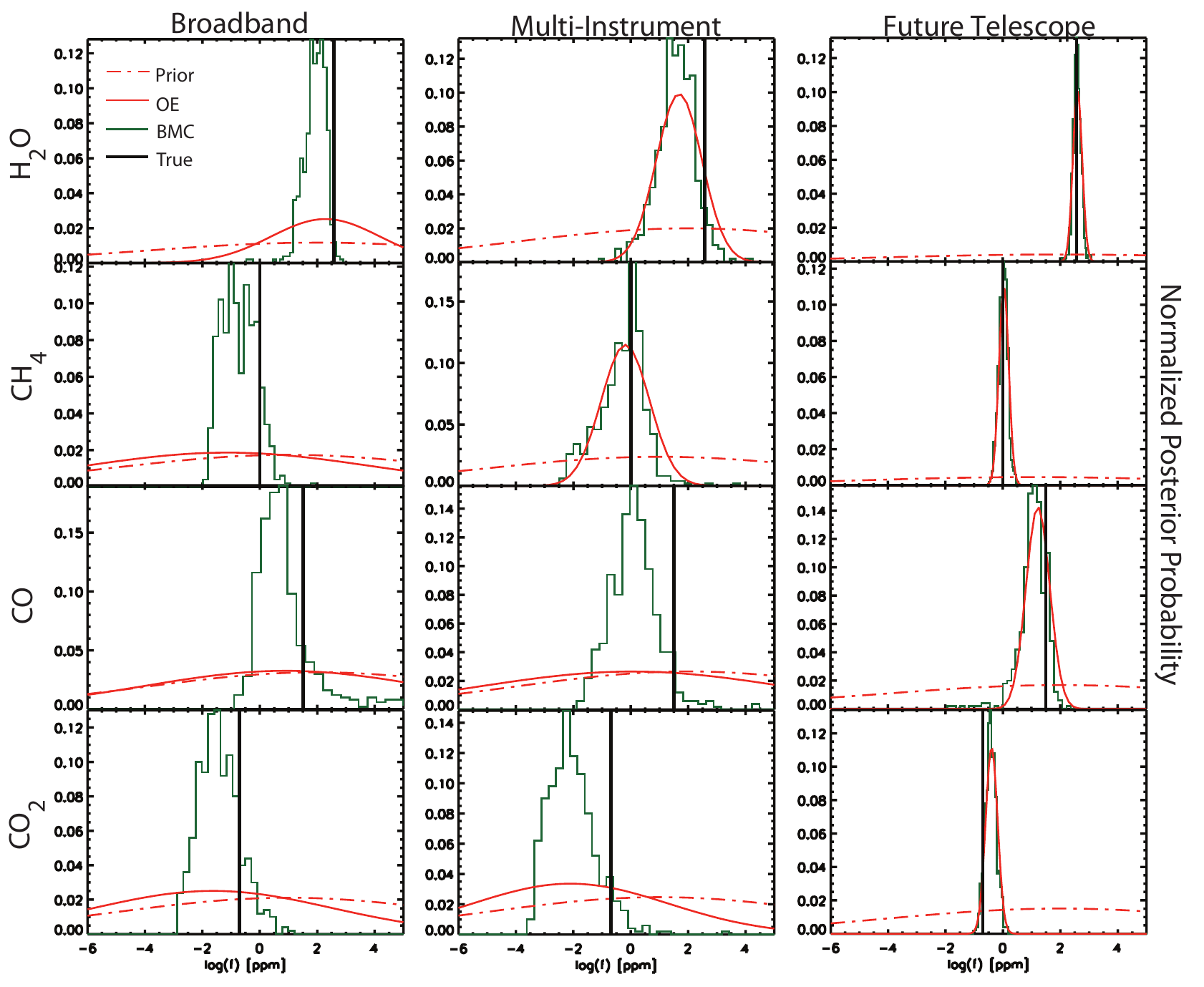}
\end{center}
\caption[Level-by-Level Retrieval Gas Uncertainties]{ \label{fig:figure11} Marginalized posterior probability distributions for each of the retrieved gases (rows) and observational scenario (columns) using the Level-by-Level temperature profile approach.   In each panel the posteriors for optimal estimation (red) and bootstrap Monte Carlo (green) are shown.  The Gaussian priors for each gas are shown with the dot dashed red curve.  The true answer is the vertical black line.   }
\end{figure*}

Figure \ref{fig:figure10} shows the spectral fits as a result of OE and BMC using the Level-by-Level temperature approach, similar to Figure \ref{fig:figure5}.   Figure \ref{fig:figure11} shows the marginalized gas posteriors.  We find the gas posteriors and the agreement amongst the retrieval techniques are very similar to those derived in Figure \ref{fig:figure6} using the parameterized temperature profile.   This is somewhat surprising given the extremely different temperature profile retrieval approaches.  This suggests that the gas abundances can be properly and consistently retrieved regardless of the temperature profile assumptions.  We could, however, imagine a contrived example in which the true temperature profile is so wildly different from what can be reasonably approximated with the parameterization,  that the two approaches would yield differing gas posteriors.
 
\begin{figure*}[h]
\begin{center}
\includegraphics[height=4.5in,width=!, angle=0]{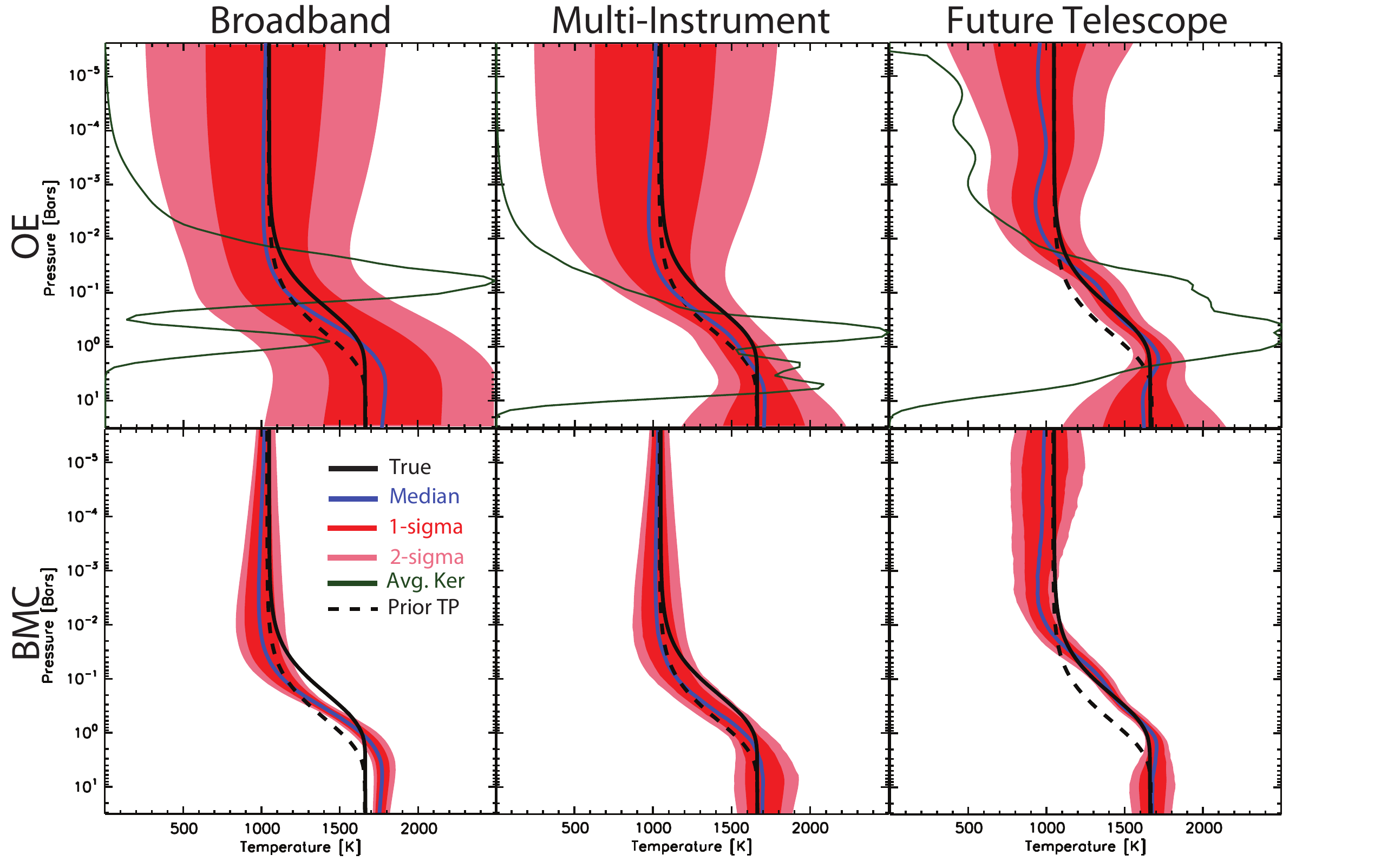}
\end{center}
\caption[Level-by-Level Temperature Profile Uncertainties]{ \label{fig:figure12}  Temperature profile posteriors using the Level-by-Level temperature profile approach for each observational scenario (columns) and two of the retrieval technique (rows).   The solid black curve in each panel is the true temperature profile constructed with equations \ref{eq:TP_par}-\ref{eq:tau} and the parameters in Table \ref{tab:table1}.  The blue curve is the median temperature profile.  The dashed black curve is the priortemperature profile constructed from $\bf x_{a}$, as in Figure \ref{fig:figure4}. The prior widths for each level (not shown) are $\pm$400 K.  The dark and light red regions are the 1- and 2$\sigma$ ($68\%$ and $95\%$) uncertainties, respectively.  The green curve is the averaging kernel profile for temperature.  The atmospheric regions over which this is a maximum is where we can retrieve temperature information with less dependence on the prior (see text).      }
\end{figure*}

Figure \ref{fig:figure12} shows temperature profile posteriors under the Level-by-Level temperature profile assumption.   For the broadband scenario,  the uncertainties more or less do not improve much beyond the prior.   The greatest gain in improvement is over the region spanned by the averaging kernel (green curve).  The uncertainty reduces from the prior uncertainty of $\pm$400 K to $\pm$265 K at 100 mbars.    The BMC approach using the Level-by-Level temperature profile produces a much smaller error than the OE approach, and at some levels the 2$\sigma$ uncertainties do not even capture the true state.  Again, this because the BMC is only  able to characterize a highly localized region around the OE best fit.  

The reduction in temperature uncertainty due to the addition of the WFC3 and ground-based photometry data is more apparent with the Level-by-Level approach than with the parameterized approach.  The uncertainties in temperature at 100 mbars are reduced to $\pm$177 K, though smaller uncertainties are achieved at deeper levels due to the addition of the WFC3 data which probe deeper atmospheric levels.  This is why the averaging kernel profile peaks at a deeper level.   The BMC results show a larger uncertainty than they do in the broadband observational scenario but still greatly underestimate the profile spread relative to optimal estimation.  

 The future spaceborne telescope observations reduce the temperature uncertainties to $\pm$70K, a factor of nearly four better than what can be done with the broadband observations.   Outside of the region spanned by the averaging kernel uncertainties relax back to the prior widths.   As before, in both cases the BMC approach underestimates the temperature uncertainties relative to the OE derived uncertainties.   The uncertainties in temperature derived using the Level-by-Level temperatures are a factor of two larger than those derived with the parameterized temperature profile.  This is because the retrievals with the parameterization only allow temperature profiles that conform to radiative equilibrium whereas the Level-by-Level retrievals can allow for a wider range of possibilities that do not necessarily have to conform to this constraint.  

\begin{figure*}[h]
\begin{center}
\includegraphics[height=3.5in,width=!, angle=0]{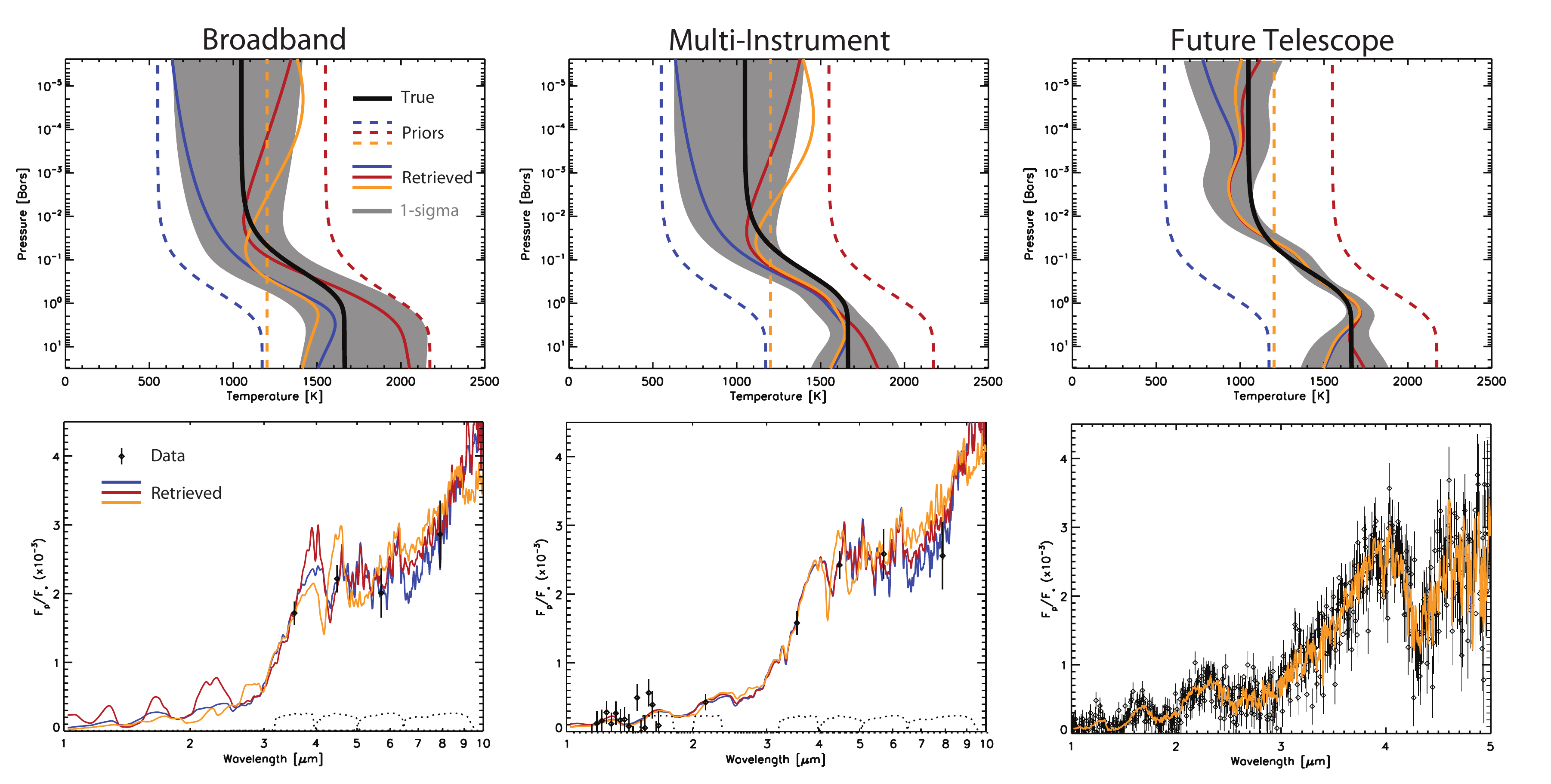}
\end{center}
\caption[Effects of Different Temperature Profile Priors]{ \label{fig:figure13} The effect of three different Level-by-Level temperature profile priors on the retrieved temperatures (top) and spectra (bottom).  The different temperature profile priors are shown as the colored dashed curves.  The prior widths (not shown) at each level are $\pm$400 K, similar to those in Figure  \ref{fig:figure12}.   The resultant retrieved profiles are shown as the solid colored curves.  The thick black curve is the true temperature profile.  The solid grey region is the 1$\sigma$ confidence interval from the retrievals in Figure \ref{fig:figure12}.   Note how the retrieved profiles all converge within the 1$\sigma$ confidence interval regardless of the temperature prior.  The best agreement is in the middle atmosphere where the thermal emission weighting functions are a maximum, and hence the averaging kernel profiles from Figure \ref{fig:figure12} are also a maximum.  The spectra in the second row illustrate the effects of the different retrieved temperature profiles of corresponding color.  There is virtually no difference in the resultant spectra for high quality data.  The dotted curves at the bottom are the broadband filter functions.       }
\end{figure*}

Another way to determine the robustness of the Level-by-Level retrieval is to explore the role of the prior temperature profile (see e.g., Lee et al. 2012).  For this, we investigate the effect of three different temperature priors (different prior profiles, $\bf x_a$ but same widths, $\bf S_a$) on the retrieved profiles and check to see if they are consistent with the estimated errors (Figure \ref{fig:figure13}).   The shaded grey region in Figure \ref{fig:figure13} is the 1$\sigma$ retrieval uncertainty from Figure \ref{fig:figure12} using the nominal prior.  Two of the other priors are the nominal profile with a $\pm$500 K offset, and the third is an isothermal profile set to the equilibrium temperature of the planet.  In all three cases we find that the retrieved profiles fall with in the 1$\sigma$ bounds of the nominal retrieved profile.  This suggests that although different temperature profile priors are used, they generally produce retrieved profiles that are statistically consistent with each other.  As the spectral quality improves, the different priors produce identically the same retrieved profiles over the atmospheric regions spanned by the thermal emission weighting functions.  Outside of this region, the profiles diverge and relax towards their respective priors with no consequence on the spectra.    This is yet another demonstration that the spectra are only sensitive to a small region of the atmosphere between a few bars up to a few mbars.  While some of these Level-by-Level profiles may not be physical, especially in the broadband observational scenario, they are a more direct reflection of the information provided by the measurements in the absence of a parameterized model.

\section{Discussion \& Conclusions}
We have developed a new statistically robust suite of exoplanet atmospheric retrieval algorithms known as CHIMERA.  This suite consists of the optimal estimation (OE, \S \ref{sec:OE} ), bootstrap Monte Carlo (BMC, \S \ref{sec:BMC}), and differential evolution Markov chain Monte Carlo (DEMC, \S \ref{sec:DEMC}) approaches and a validated forward model (\S \ref{sec:forward_model}).  We have tested each of these approaches on the dayside thermal emission spectra for a synthetic planet under a variety of observational scenarios ranging from current observations to potential future observations (\S \ref{sec:three}).   In general, we find that the three retrieval approaches produce similar posteriors when the measurement quality is good, typically when there are more observed spectral channels than retrievable parameters (Figures \ref{fig:figure6} and \ref{fig:figure7} ).   The Gaussian approximation made by optimal estimation breaks down and becomes invalid for low resolution measurements, but is perfectly valid for high resolution measurements likely to come from future spaceborne observations.  This approach also appropriately captures the correlations amongst the various parameters.   This approach is much less of a computational burden than Monte Carlo approaches and will prove useful for quick reductions of large, high-quality data sets.  The optimal estimation formalism also allows for the calculation of the averaging kernel (\S \ref{sec:par_vs_Level-by-Level}), which is a useful diagnostic to determine how much of the posterior is influenced by the prior versus the measurements.    The bootstrap Monte Carlo approach generally fails to capture the essence of the posteriors.  This is because the regeneration of synthetic measurement realizations based on the optimal estimation best fit only sample a localized region of phase space near the best fit solution.  This is especially problematic in the cases where there are fewer spectral data points than parameters.  In this scenario, even with small measurement error, there will still be many possible best-fit solutions, thus creating enormous degeneracies among the parameters.  Since the BMC is initialized with only one possible best-fit set of parameters out of many, the derived parameter uncertainties will only be representative of the uncertainties about that localized best fit.    We therefore strongly advise against the bootstrap Monte Carlo approach when the number of parameters is larger than the number of spectral data points.   In the high signal-to-noise and high spectral resolution regime, both the BMC and OE methods provide reasonable parameter uncertainties.   We have also introduced the application of differential evolution Markov chain Monte Carlo to the spectral retrieval problem and found that convergence can be obtained efficiently by using an appropriate proposal distribution based on the chain history.  This approach appears to be valid in all observational scenarios but requires many hundreds of thousands of forward model calls.     

We find that for the particular combination of gas abundances in our synthetic planet, the broadband observations provide limited constraints on the gas abundances.  For this example planet, the Spitzer photometry does a particularly poor job constraining the relative abundance of CO, with most posteriors simply reflecting their priors (\S \ref{sec:gas_retrieval_par}).  The addition of WFC3 observations provide little additional constraint on the gas abundances derived from dayside thermal emission spectra, with the exception of a slight improvement on the water abundances.   This is primarily due to the limited spectral coverage provided by the red grism on WFC3, which spans the wavelengths from 1.2-1.6 $\mu$m.  Admittedly, our choice of molecular abundances is unfair to CH$_4$, CO and CO$_2$, so our conclusions are somewhat pessimistic for these molecules.  We could imagine other planets with greater abundances of these species which would provide more spectral leverage and hence better constraints.  A wider range of compositions will be explored in future investigation.  As the measurement quality improves, the parameter uncertainties decrease and become more Gaussian.  Moderate cost future spaceborne instruments have the capability of obtaining better than order-of-magnitude constraints on gas compositions with their posteriors generally being independent of the prior.  This is typically many orders of magnitude better than current observational capabilities (Table \ref{tab:table3}).   The derived gas posteriors are also independent of whether or not a parameterized or Level-by-Level temperature profile is used (Figure \ref{fig:figure6} vs. Figure \ref{fig:figure11}).  We also find strong correlations of the water abundance, if it is abundant, with temperature as has been pointed out in other investigations (see e.g., Lee et al. 2012 for a nice example) in all observational scenarios. Correlations amongst the gas abundances themselves become more prevalent in the higher resolution future telescope scenario.

Constraining the C to O ratio of exoplanet atmospheres is very difficult due to the broad nature of some of the gas posteriors, especially CO.  In the absence of valid observational constraints the posteriors for these molecules simply reflect the priors, which produce a double-peaked distiribution with maxima at C to O ratios of 0.5 and 1 (\S \ref{sec:CtoO}, Figure \ref{fig:figure9}).  Only high quality observations from the future spaceborne telescope scenerio are independent of the double-peaked prior.  As a result, caution must be taken when interpreting C to O ratios from broad gas posteriors.

Reasonable temperature constraints could be obtained in all observational scenarios and temperature retrieval approaches, though the bootstrap Monte Carlo approach again fails to fully capture the posterior (\S \ref{sec:TP_retrieval_par}, Figure \ref{fig:figure8}).  The temperature profiles and corresponding uncertainties can only be trusted for the region over which the thermal emission contribution functions peak, typically between a few bars and a few mbars (Figure \ref{fig:figure2}).  Outside of this window, the temperature profiles are strongly affected by their priors.   The Level-by-Level temperature profile approach overestimates the temperature uncertainties compared with the parameterization due to the allowance of more profiles (\S \ref{sec:Level-by-Level}).  These Level-by-Level profiles can be unphysical but are more reflective of the measurements without imposing preconceived notions of how the physical structure of the atmosphere should behave.  While this approach produces statistically consistent profiles in low quality observational scenarios,  we would still recommend using a parameterization for said cases.  However, for high quality spectra the Level-by-Level approach is recommended given its slightly more pessimistic temperature uncertainties and its non-dependence on a particular parameterization.   

In a follow up investigation we will use CHIMERA to perform a uniform analysis of an ensemble of secondary eclipse spectra.    Such a study will allow us to determine the biases introduced by the choice of fitting method for individual planets and to derive a uniform set of relative abundances and temperatures for these planets that can be reliably inter-compared and trends identified.  This kind of uniform analysis has the potential to provide invaluable insights into exoplanetary atmospheric processes and formation environments.

\section{Acknowledgements}		
We thank Jaimin Lee and Leigh Fletcher for their willingness to compare radiative transfer codes.   We also thank John Johnson and Jonathan Fortney for useful conversations.     We thank members of Yuk YungÕs group for useful comments. This research was supported in part by an NAI Virtual Planetary Laboratory grant from the University of Washington to the Jet Propulsion Laboratory and California Institute of Technology. Part of the research described here was carried out at the Jet Propulsion Laboratory, California Institute of Technology, under a contract with the National Aeronautics and Space Administration. YLY was supported in part by NASA NNX09AB72G grant to the California Institute of Technology.

\begin{table}[h]
\centering
\caption{\label{tab:table1} Parameters used to generate the fictitious model atmosphere and spectrum.  $R_p$ is the planet radius in units of Jupiter radii, $R_{star}$ is the stellar radius in units of solar radii, $T_{star}$ is the stellar effective temperature, $a$ is the semimajor axis, $T_{int}$ is the internal heat flux temperature of the planet, $g$ is the planetary surface gravity.  $\gamma_{v1}$, $\gamma_{v2}$, $\kappa_{IR}$, $\alpha$, and $\beta$ are the parameters that control the shape of the radiative equilibrium temperature profile.  The $f_i$'s are the constant-with-altitude volume mixing ratios for each species in parts-per-million (ppm).  }
\begin{tabular}{cc}
\cline{1-2}
Parameter & Value \\
\hline
$R_p (R_J)$&1.138\\
$R_{star} (R_{sun})$ & 0.756\\
$T_{star} (K)$ & 5040\\
$a (AU)$ & 0.031\\
$T_{int} (K)$ & 100 \\
$log(g) (cm s^{-2})$ & 3.341\\
$\gamma_{v1}$ & $1.58\times 10^{-1} $ \\
$\gamma_{v2}$& $1.58\times 10^{-1} $  \\
$\kappa_{IR} (cm^2 g^{-1})$ & $3\times 10^{-2} $  \\
$\alpha$   &  0.5  \\
$\beta$ &  1.0 \\
$f_{H2}$ (ppm) & $8.5\times 10^5$ \\
$f_{He}$ (ppm) & $1.5\times 10^5$ \\
$f_{H2O}$ (ppm) & 370 \\
$f_{CH4}$ (ppm) &  1 \\
$f_{CO}$ (ppm)&  31.6 \\
$f_{CO2}$ (ppm) &  0.2 \\
\hline
\end{tabular}
\end{table}

\begin{table}[h]
\centering
\caption{\label{tab:table2}Gaussian prior parameter values and widths.  The true state is the same as in Table \ref{tab:table1} but in logarithmic units for some of the parameters.  The mixing ratios of each gas, $f_{k}$, are in ppm.  The infrared opacity, $\kappa_{IR}$, has units of $cm^2 g^{-1}.$  $\gamma_{v1}$,  $\gamma_{v2}$, $\alpha$, and $\beta$ are all unit-less.  Note that we retrieve the log of all values except $\alpha$ and $\beta$.}
\begin{tabular}{cccc}
\cline{1-4}
Parameter & True & Prior State ($x_{a,i}$)  & Prior Width ($\sqrt(S_{a,ii})$) \\
\hline
$log(\gamma_{v1})$& -0.8  & -0.9 & 1 \\
$log(\gamma_{v2})$& -0.8 & -0.7 & 1\\
$log(\kappa_{IR})$&  -1.52& -2.0 &  0.5 \\
$\alpha$  & 0.5 &0.5 & 0.05  \\
$\beta$  & 1 &1 & 0.25  \\
$log(f_{H2O})$& 2.568 & 2 & 6 \\
$log(f_{CH4})$& 0.0 & 1 &  6\\
$log(f_{CO})$& 2.663 & 2&  6\\
$log(f_{CO2})$& -0.70 & 1&  6\\
\hline
\end{tabular}
\end{table}

\begin{table}[h]
\centering
\caption{\label{tab:table3}Numerical summary of the retrieval results for several parameters as derived from each retrieval technique and observational scenario.  For each parameter and each observational scenario we show the true value, the 1$\sigma$ (68$\%$ confidence interval) marginalized prior uncertainties, and the 1$\sigma$ marginalized uncertainties derived from optimal estimation (OE), bootstrap Monte Carlo (BMC), and differential evolution Markov chain Monte Carlo (DEMC) as well as the averaging kernal element for that paramter (A) .  The gas abundances, $f_{i}$, are given in terms of volume mixing ratio.  We also show a representative temperature (100 mbars temperature) and the C-to-O ratio.  This table is layed out so that for a given parameter easy comparisons in either the observational scenario (left-right) or the retrieval techniques (top-bottom) can be made.    }
\begin{tabular}{llccc}
\hline
\cline{1-2}
Paramter    &   & Broadband & Multi-Instrument & Future Telescope \\
\hline
$f_{H2O}$     & True: & $3.70\times 10^{-04}      $&$    3.70\times 10^{-04}  $&$ 3.70\times 10^{-04}  $    \\
                        & Prior:  &  $4.94\times 10^{-10}-3.92\times 10^{-03}    $&$ 4.94\times 10^{-10}-3.92\times 10^{-03}    $&$  4.94\times 10^{-10}-3.92\times 10^{-03} $ \\
                        & OE:   & $1.25\times 10^{-06}-7.82\times 10^{-03}     $&$ 3.68\times 10^{-06}-7.03\times 10^{-04}   $&$  2.31\times 10^{-04}-4.56\times 10^{-04} $ \\
                        & BMC: &  $5.74\times 10^{-05}-2.68\times 10^{-04}   $&$  1.18\times 10^{-05}-2.45 \times 10^{-04}  $ &$  2.44\times 10^{-04}-4.03\times 10^{-04} $ \\
                        & DEMC: & $1.88\times 10^{-07}-1.24\times 10^{-03}$&$  7.02 \times 10^{-06}- 1.38\times 10^{-03}   $ &$ 2.83\times 10^{-04}-4.97\times 10^{-04}  $  \\
                        & A: &  0.872 &  0.983  &  0.999  \\
\hline                        
$f_{CH4}$     & True: &$  1.00\times 10^{-06}     $&$  1.00\times 10^{-06}    $&$ 1.00\times 10^{-06}  $    \\
                        & Prior: &$  2.63\times 10^{-10}-2.55\times 10^{-03}     $&$   2.63\times 10^{-10}-2.55\times 10^{-03}  $&$ 2.63\times 10^{-10}-2.55\times 10^{-03}   $\\
                        & OE:  &$   8.89\times 10^{-11}-5.73\times 10^{-04}    $&$   4.54\times 10^{-09}- 3.92\times 10^{-05}  $&$  7.50\times 10^{-07}-1.54\times 10^{-06}  $\\
                        & BMC: &$   6.40\times 10^{-08}-1.64\times 10^{-06}  $&$  9.07\times 10^{-08}- 3.89\times 10^{-06}     $&$  7.62\times 10^{-07}-1.46\times 10^{-06}  $\\
                        & DEMC:&$ 1.87\times 10^{-11}-1.98\times 10^{-07} $&$   4.45\times 10^{-11}- 1.94\times 10^{-06}  $&$  7.31\times 10^{-07}-1.52\times 10^{-06}  $ \\
                        & A:& 0.259  & 0.979 & 0.999 \\
\hline       
$f_{CO}$     & True: &$  3.16\times 10^{-05}     $&$  3.16\times 10^{-05}    $&$   3.16\times 10^{-05} $    \\
                        & Prior: &$ 4.94\times 10^{-10}-3.92\times 10^{-03}      $&$ 4.94\times 10^{-10}-3.92\times 10^{-03}    $&$ 4.94\times 10^{-10}-3.92\times 10^{-03}   $\\
                        & OE:  &$  4.97\times 10^{-10}-2.43\times 10^{-03}     $&$  7.40\times 10^{-10}- 4.13\times 10^{-03}    $&$ 5.87\times 10^{-06}-3.13\times 10^{-05}   $\\
                        & BMC: &$  2.05\times 10^{-06}-6.10\times 10^{-05}   $&$  3.00\times 10^{-07}- 1.81\times 10^{-05}    $&$4.37\times 10^{-06}-2.76\times 10^{-05}    $\\
                        & DEMC:&$ 2.08\times 10^{-10}-5.35\times 10^{-04} $&$   8.93\times 10^{-11}- 7.06\times 10^{-05}    $&$  3.69\times 10^{-06}-2.67\times 10^{-05}  $ \\
                        & A:&  0.316 & 0.176  &0.996\\
\hline                                                                   
$f_{CO2}$     & True: &$  2.00\times 10^{-07}     $&$   2.00\times 10^{-07}   $&$ 2.00\times 10^{-07} $    \\
                        & Prior: &$ 2.63\times 10^{-10}-2.55\times 10^{-03}      $&$ 2.63\times 10^{-10}-2.55\times 10^{-03}    $&$ 2.63\times 10^{-10}-2.55\times 10^{-03}    $\\
                        & OE:  &$  7.73\times 10^{-11}-1.78\times 10^{-04}     $&$ 5.07\times 10^{-09}- 4.25\times 10^{-03}    $&$  1.94\times 10^{-07}-4.82\times 10^{-07}  $\\
                        & BMC: &$  9.44\times 10^{-09}-3.61\times 10^{-07}   $&$  1.64\times 10^{-09}- 1.10\times 10^{-07}     $&$ 2.03\times 10^{-07}-4.29\times 10^{-07}   $\\
                        & DEMC:&$2.21\times 10^{-11}-9.01\times 10^{-07}  $&$ 2.33\times 10^{-11}- 7.36\times 10^{-07}   $&$  2.29\times 10^{-07}-5.02\times 10^{-07} $ \\
                        & A:& 0.508 &  0.689 & 0.999 \\
\hline                                                                   
$T_{100mb} [K]$     & True: &$  1313     $&$   1313   $&$  1313 $    \\
                        & Prior: &$  876-1503     $&$ 876-1503    $&$  876-1503  $\\
                        & OE:  &$   932-1358    $&$  1075-1274  $&$  1267-1340  $\\
                        & BMC: &$  1150-1249   $&$ 1117-1284    $&$  1278-1327  $\\
                        & DEMC:&$ 1048-1355  $&$ 1135-1373    $&$ 1294-1348   $ \\
\hline                                                                   
$C/O$     & True: &$   8.00\times 10^{-2}    $&$ 8.00\times 10^{-2}      $&$ 8.00\times 10^{-2}  $    \\
                        & Prior: &$   3.82\times 10^{-02}-8.00    $&$ 3.82\times 10^{-02}-8.00  $&$ 3.82\times 10^{-02}-8.00   $\\
                        & OE:  &$   2.07\times 10^{-03}-1.00    $&$  6.26\times 10^{-03}- 0.938     $&$  2.25\times 10^{-2}- 8.93\times 10^{-2} $\\
                        & BMC: &$   1.60\times 10^{-02}-0.32  $&$ 1.29\times 10^{-02}- 0.427     $&$ 1.81\times 10^{-02}-8.09\times 10^{-2}   $\\
                        & DEMC:&$ 3.75\times 10^{-04}-0.970 $&$   1.94\times 10^{-04}-0.720   $&$  1.33\times 10^{-02}-7.00\times 10^{-2} $ \\
\\                                                                                                                                    
\hline
\end{tabular}
\end{table}

\end{document}